\documentclass[12pt]{article}
\usepackage[utf8]{inputenc}
\usepackage[english]{babel}
\pdfoutput=1 %arXiv fully supports and automatically recognizes PDFLaTeX. You can ensure pdflatex processing by setting the flag \pdfoutput=1 within the first 5 lines of the preamble of the main .tex file. You should not need any other special flag.

\usepackage{graphicx}% Include figure files
\usepackage{bm}% bold math
\usepackage{latexsym}
\usepackage{dcolumn}
\usepackage{amsmath}
\usepackage{epsfig,amssymb,euscript,mathrsfs}
\usepackage{array,calc,epsfig}

\usepackage{amsfonts}
\usepackage{amsthm}
\usepackage{slashed}
\usepackage{caption}
\usepackage{pdfpages}
\usepackage{braket}
\usepackage{rotating}
\usepackage{bbm}

\usepackage{tikz-cd}

\usepackage[noadjust]{cite}
\usepackage{chngcntr}
\usepackage[pdfpagelabels]{hyperref}
\usepackage{color}

\usepackage{booktabs}
\usepackage{comment}
\usepackage{cases} %per scrivere sistemi di equazioni labellate
\usepackage{empheq}
\usepackage{cancel}

\usepackage{xfrac} %per frazioni storte (\sfrac{}{})
\usepackage{float} %per \begin{figure}[H]
\usepackage{subfig}

\newcommand{\numberset}{\mathbb}
\newcommand{\N}{\numberset{N}}
\newcommand{\R}{\numberset{R}}

\newcommand{\Z}{\numberset{Z}}

\newcommand{\p}{\partial}

\newcommand{\T}{\mathcal{T}}

\newcommand{\A}{\mathcal{A}}

\renewcommand{\d}{\mathrm{d}}

\renewcommand{\Im}{\text{Im}}
\renewcommand{\ker}{\text{Ker}}

\newcommand{\spin}{\text{Spin}}
\newcommand{\SO}{\text{SO}}
\newcommand{\U}{\text{U}}

\newcommand{\aut}{\text{Aut}}
\renewcommand{\hom}{\text{Hom}}

\def\a{\alpha} 
\def\b{\beta} 
\def\g{\gamma} 
\def\G{\Gamma}

\def\e{\epsilon}

\def\l{\lambda} 
\def\L{\Lambda}

\def\s{\sigma} 
\def\S{\Sigma}

\def\w{\omega} 
\def\W{\Omega} 
 
\def\th{\theta}
\def\wt{\widetilde} 
\def\wh{\widehat}

\numberwithin{equation}{section}

\let\oldlgraf\{ %rinomino la parentesi graffa con un altro nome, così posso usare \{ per la ridefinzione del comando
\renewcommand{\{}{\left \oldlgraf}

\let\oldrgraf\}
\renewcommand{\}}{\right \oldrgraf}

\newlength\dlf  % Define a new measure, dlf

\title{On gauging Abelian extensions of finite and U(1) groups}
\author{}
\date{}

\begin{document}

\begin{titlepage} 
\begin{center} 
\vspace{1cm}
{\LARGE On gauging Abelian extensions of finite and U(1) groups}
\vspace{1cm}

Riccardo Villa${}^{a}$

\vspace{1cm}
{\em ${}^{a}$INFN, Sezione di Firenze,\\
  Via G. Sansone 1, 50019 Sesto Fiorentino - Firenze, Italy}\\

\end{center}

\vspace{1cm}

\begin{abstract}
  We consider Abelian extensions of global symmetries of the form $A \to G \to K$, with $A$ finite. For a quantum field theory $\mathcal{T}$ with symmetry $G$, we compare gauging $G$ directly with gauging first $A$ and then $K$, and show that for finite Abelian groups and for $K \simeq \U(1)$ the two procedures are equivalent as expected, $\mathcal{T}/G \simeq \mathcal{T}/A/K$. In the continuous case $K=\U(1)$, after gauging the full extension, the dual symmetry $\widehat{\mathbb{Z}}_q^{(d-2)}$ fits into an extension characterizing the topological data of the magnetic $\U(1)_m^{(d-3)}$ symmetry. This is better described using differential cohomology. 
\end{abstract} 
 
\vfill 
\end{titlepage} 
\newpage

\tableofcontents
\newpage

\section{Introduction}

In these notes we consider Abelian extensions of finite and $\U(1)$ groups of the form
\begin{equation}
\label{intro grp ext}
   A\to G \to K.
\end{equation}
All groups are Abelian and $A$ is finite. The group $G$ is assumed to be a global symmetry of a quantum field theory $\T$, and we ask what happens when $G$ is gauged in two steps: first gauging $A$ and then gauging the remaining symmetry $K$, i.e. $\T \to \T/A\to \T/A/K$. This procedure should be equivalent to gauging $G$ directly, so that $\T/G \simeq \T/A/K$. We prove this statement in general (without discrete torsions) for finite Abelian groups and for the case $K \simeq \U(1)$. The latter reduces to the finite-group case when the gauging of $\U(1)$ involves only flat connections.

It is well known that gauging $A$ in \eqref{intro grp ext} produces a product symmetry $\wh A^{(d-2)} \times K$ with a mixed anomaly \cite{tachikawaGaugeFiniteGroups}, where $\wh A \coloneqq \hom(A,\R/\Z)$. In two dimensions, finite symmetries are completely understood in terms of fusion categories (see \cite{bhardwajTachikawa2017} for a review) and the equivalence $\T/G\simeq \T/A/K$ has been shown in \cite{ThomasNonInvSymRepsII}, even when the groups involved are non-Abelian and the dual symmetries are typically non-invertible. In the present work we restrict to the Abelian case, so that the dual symmetries remain invertible higher-form symmetries \cite{gaiotto2015generalized}, and extend the discussion to arbitrary spacetime dimension and, in particular, to the continuous case $K \simeq \U(1)$.

The paper is organized as follows. Section \ref{sec grp ext} is a review of the theory of group extensions, with focus on the classification of central extensions, and it serves also to set the notation.  It may be skipped by readers already familiar with the subject. In Section \ref{App GrpExt} we analyze the effect of gauging \eqref{intro grp ext} when all groups are finite and show in arbitrary dimension that $\T/G \simeq \T/A/K$. The discussion extends straightforwardly to extensions of higher-form symmetries of the same degree. Section \ref{U(1) extensions sec} treats the continuous case $K=\U(1)$ with $A=\Z_q$. This is perhaps the most non-trivial part of these notes: after gauging the full extension, the dual symmetry of $A$, namely $\wh A^{(d-2)} = \wh \Z_q^{(d-2)}$, becomes part of the topological data of the magnetic symmetry $\U(1)_m^{(d-3)}$. More precisely, the background field for $\wh \Z_q^{(d-2)}$ gives the mod-$q$ reduction of the generalized first Chern class of $\U(1)_m^{(d-3)}$. This result implies a dual extension of $\wh \Z_q^{(d-2)}$ by a magnetic symmetry $\wt \U(1)_m^{(d-3)}$ and it can be described naturally using a differential cohomology approach. Finally, in Section \ref{sec symfrac} we make a brief comment on symmetry fractionalization. Appendix \ref{app sym frac} contains some other facts about symmetry fractionalization to complete the discussion of Section \ref{sec symfrac}.

\section{Generalities on group extensions}
\label{sec grp ext}

This section is an overview of the topic of group (and, in particular, central) extensions.\footnote{In the first part we strictly follow the notes by G.W. Moore at \url{https://www.physics.rutgers.edu/~gmoore/PiTP-LecturesA.pdf}.}

A group extension $G$ of $K$ by $A$ is given by the following short exact sequence 
\begin{equation}
\label{general grp ext}
   1 \to A \xrightarrow{\iota} G \xrightarrow{\pi} K \to 1.
\end{equation}
$\iota$ is an injective homomorphism, $\pi$ is a surjective homomorphism, with $\Im(\iota)=\ker (\pi)$. $\iota(A) \simeq A$ is a normal subgroup of $G$ ($gag^{-1}\in A$, $\forall g\in G,a\in A$), being the kernel of $\pi$, and $G/A \simeq K$ (which is a group, since $A$ is normal). The extension \eqref{general grp ext} can be thought of as a fiber bundle over $K$ with fiber $A$. It is always possible to choose a section  $s: K \to G$ such that $\pi(s(k))=k$. To any such section, one can associate a map $\rho: K \to \aut(A)$, which sends $k\mapsto\rho_k$ defined as\footnote{The right hand side is still in $A$ since $A$ is normal. Moreover, to be precise, it will be $\iota(\rho_k(a))=s(k)\iota(a)s(k)^{-1}$, which is well-defined for $\rho$ since $\iota$ is injective. Here and in the following we often identify $A$ with $\iota(A)$ dropping the injective $\iota$.}
\begin{equation}
    \rho_k(a)=s(k)as(k)^{-1}.
\end{equation}
One can check that $\rho_k$ is indeed an automorphism of $A$. Usually $s$ is not a homomorphism. When $s$ is a homomorphism, the sequence \eqref{general grp ext} splits, and $G \simeq A \rtimes_\rho K$. This can be seen using the isomorphism $\iota(a)s(k)\mapsto(a,k)$, with $(a_1,k_1)\cdot(a_2,k_2) = (a_1\rho_{k_1}(a_2),k_1k_2)$.

When $A$ in \eqref{general grp ext} is Abelian and $\iota(A)\subset Z(G)$, the extension is called a central extension.\footnote{They are ubiquitous in physics. A common example is $\Z_2\to \spin (d)\to \SO(d)$. More generally, projective representations of a group $G$ are connected to central extensions by $\U(1)$, classified by $H^2(G;\U(1))$.} Isomorphism classes\footnote{Take two extensions $G$ and $G'$ of the form \eqref{general grp ext}. If there is an isomorphism $\phi: G\to G'$ and the diagram that they form commutes, then the two extensions are isomorphic.} of central extensions of a group $K$ by $A$ are classified (namely, there is a bijective correspondence) by $H^2_{grp}(K;A) \simeq H^2(BK;A)$, with $BK$ the classifying space of $K$ bundles. Therefore, to any such extension, one can associate a class $[\a]\in H^2(BK;A)$. If $[\a]=0$, the extension splits and $G\simeq A\times K$ (notice that here $\rho_k(a)=a$ is trivial since $A$ is central, so the semidirect product of before is a direct product).

Let us see how the cohomological data arise. In general, a sequence \eqref{general grp ext} splits if there is a section $s$ that is a homomorphism. It is possible to measure how $s$ fails to be a homomorphism by considering
\begin{equation}
    \e = s(k_1)s(k_2)s(k_1k_2)^{-1}.
\end{equation}
Notice that $\e \in\ker(\pi) = \Im (\iota)$. It is therefore possible to introduce a function $\a_s : K\times K \to A$ such that $\a_s(k_1,k_2)= \e$, or, in other words,
\begin{equation}
    s(k_1)s(k_2)=s(k_1k_2)\a (k_1,k_2).
\end{equation}
The failure of $s$ to be a homomorphism is thus measured by a 2-cochain $\a \in C^2(K;A)$. By considering the product $s(k_1)s(k_2)s(k_3)$ it is straightforward to show that 
\begin{equation}
    \a_s(k_1,k_2) \a_2(k_1k_2,k_3)=\a_s(k_2,k_3)\a_s(k_1,k_2k_3),
\end{equation}
i.e. $\a_s$ is a cocycle $\a_s\in Z^2(K;A)$, $\d \a_s=0$. This object depends on the choice of $s$. However, if we consider a different section $\wh s(k)= s(k)\s(k)$, with a function $\s : K \to A$, then $\a_s$ and $\a_{\wh s}$ just differ by a coboundary,
\begin{equation}
    \a_{\wh s} (k_1,k_2) = \a_s (k_1,k_2)\s(k_1)\s(k_2)\s(k_1k_2)^{-1}= \a_s(k_1,k_2) (\d \s)(k_1,k_2).
\end{equation}
Therefore, to any central extension we can associate unambiguously a class $[\a] \in H^2(BK;A)$, independent from the choice of $s$. Finally, if two extensions are isomorphic, $\phi: G \to G'$, for any section $s$ there is a natural section $s'= \phi \circ s$. It trivially follows that $s'(k_1)s'(k_2)= s'(k_1k_2)\a_s(k_1,k_2)$ for the same cocyle $\a_s$. So, for any isomorphism class of a central extension, we can construct $[\a]$ uniquely.

On the converse, from any element $[\a]\in H^2(BK;A)$ it is possible to construct a central extension \eqref{general grp ext}. Choosing a representative $\a\in[\a]$, we construct a group $G$ that is $A\times K$ as a set, but with a twisted product rule\footnote{It is easy to check that $G =\{(a,k)\}$ is an extension of $K$ by $A$ by considering the inclusion $\iota: A\hookrightarrow G$ and the natural projection on the right component $\pi: G \to K$. $A$ is a normal subgroup given the product rule \eqref{grp ext product}.}
\begin{equation}
\label{grp ext product}
     (a_1,k_1) \cdot (a_2,k_2) = (a_1a_2\a(k_1,k_2),k_1 k_2).
\end{equation}
The cocycle condition ensures associativity. If we choose two different representatives of $[\a]$ we obtain two isomorphic extensions. Indeed, let $G$ be defined by \eqref{grp ext product} and $G'$ the same but with $\a'= \a \d \s$ for some coboundary $\d \s$. Then there is an isomorphism $\phi: G\to G'$ defined by $(a,k)\mapsto (a \s(k)^{-1}, k)$. This point of view gives also a very concrete way to think about central extensions \eqref{general grp ext}: $G$ is a direct product set with twisted product rule \eqref{grp ext product}.

Notice that even when $K$ and $A$ are both Abelian, $G$ can be a non-Abelian group, as follows from \eqref{grp ext product}. $G$ is Abelian when $\a$ is a symmetric cocycle, i.e. $\a(k_1,k_2)=\a(k_2,k_1)$ for every $k_1$, $k_2 \in K$. The set of such Abelian extensions is thus $H^2_{\text{sym}}(BK;A) \subset H^2(BK;A)$. In the main text we consider  Abelian group extensions of this kind.

\paragraph{Continuous groups.} We have proven that central extensions like \eqref{general grp ext} are classified by $H^2(BK;A)$. However, we have actually used the identification $H^2_{grp}(K;A) \simeq H^2(BK;A)$ and used the group cohomology notation all around. This is strictly fine for $K$ discrete, but when $K$ is a continuous group (e.g. a Lie group) the situation is more tricky and the group cohomology for such objects is complicated, since one would like to require that all maps and differentials are continuous. It turns out that that there is a sensible way to do so and define a similar suitable differentiable group cohomology for Lie groups such that $H^2_{grp}(K;A) \simeq H^2(BK;A)$. See Appendix A of \cite{kapustinSopenko2025}. However, for our purpose, we really care about the result $H^2(BK;A)$, so we will now give an argument directly for this classification without recurring to group cohomology. This has the benefit of being homogeneous for both discrete and continuous groups, but it is certainly much more abstract than the previous discussion using group cohomology.

We first show that there is a bijection between central extensions like \eqref{general grp ext} and the corresponding fibration for the classifying spaces $BA  \to BG\to BK$. Given \eqref{general grp ext}, it is possible to proceed as follows. For any group $G$ and a contractible space $EG$ with a free $G$-action, the classifying space is $BG \simeq EG / G$. Now, the group $G$ acts on $EK$ via the projection $\pi$ in \eqref{general grp ext}.\footnote{Explicitly, for $g \in G$ and $p \in EK$, the action is $p \cdot g = p \cdot \pi(g)$. Note that the subgroup $A = \ker(\pi)$ acts trivially on $EK$.} We can thus define a diagonal free action of $G$ on $EG \times EK$, $(p_G, p_K) \cdot g = (p_G \cdot g, p_K \cdot \pi(g))$, such that $   BG \simeq (EG \times EK) / G$. The projection $EG \times EK \to EK$ gives us a map $\wt \pi: (EG \times EK) / G \simeq BG \to EK/G \simeq BK$, where $EK/G \simeq EK/K$, since $G$ acts on $EK$ only through its quotient $K \simeq G/A$. The fiber of this map $\wt \pi^{-1}([p_K])$, for $[p_K]\in BK$, is given by the equivalence classes $([p_G, p_K])$ that does not move $p_K$: these are the elements of $G$ that map trivially to $K$, i.e. the subgroup $A$. Therefore, the fiber is $EG/A \simeq BA$ (notice that $A$ acts freely on $EG$ via the injective inclusion $\iota$ of \eqref{general grp ext}). So, starting from the central extension \eqref{general grp ext}, we obtained the fiber bundle $BA \to BG \to BK$.

Conversely, starting from $BA \to BG \to BK$, we can recover \eqref{general grp ext} as follows. First, we can extract the group $G$ from its classifying space $BG$ using path lifting. Consider a loop $\g$ at a fixed basepoint $ p \in BG$ (so $\g: [0,1]\to BG$ with $\g(0)=\g(1)=p$) and fix a single point $\wt p \in EG$ such that $\wt \pi(\wt p) = p$. By the path lifting property \cite{HatcherAT}, the loop $\g$ lifts to a unique path $\wt \g$ in $EG$, starting at $\wt p$ and ending at $\wt p \cdot g$ for some $g \in G$, since the fiber is precisely the orbit of $\wt p$ under the action of $G$. In this way we recover $G$ from the space of based loops in $BG$, assigning to each loop $\g$ the element $g$ of $\wt p \cdot g$. We then simply apply this property to $BA \to BG \to BK$. Take a loop $\gamma_G$ in $BG$, whose lifting to $EG$ defines an element $g \in G$. Given the projection $\wt \pi: BG \to BK$ we obtain a loop $\gamma_K = \wt\pi \circ \gamma_G$ in $BK$, which again gives an element $k \in K$. This gives a homomorphism $\pi: G \to K$ (so that $g \mapsto k$). The kernel of this map consists of elements $g \in G$ that map to the identity in $K$, which corresponds to the constant trivial loop in $BK$. Therefore, the kernel consists of loops in $BG$ that are projected to the basepoint in $BK$, i.e. the loops that lie entirely within the fiber $BA$, which, again, reconstruct $A$ by lifting to $EA$. So $A \simeq \ker(\pi)$ and we recover the sequence \eqref{general grp ext}.\footnote{This argument on the correspondence between a group extension and the fibration for the classifying spaces has been discussed with Gemini 3.1 Pro. It proposed that $(EG \times EK)/G$ is indeed a model for $BG$ and the explicit path lifting construction of $G$ from $BG$ that reconstructs the group extension.}

The conclusion is that to classify central extensions \eqref{general grp ext} we could look at bundles $BA \to BG \to BK$, for $BA = K(A,1)$ an Eilenberg-MacLane space, since we restrict to finite Abelian $A$.\footnote{$K(A,n)$ is defined as a space with $\pi_n\simeq A$ and $\pi_i$ trivial for all $i\neq n$ \cite{HatcherAT}. For a finite Abelian group $A$, $K(A,1)$ has the role of $BA$.} For a group $G$, one can always find $EG$ that is also a group (with a $G$ subgroup) and, furthermore, if $G=A$ is Abelian, also $BA$ is an Abelian group \cite{Steenrod1968}. So we can choose a specific $K(A,1)$ which is an Abelian group.\footnote{Generically, $K(A,n)$ are $H$-groups, i.e. spaces where the group properties hold up to homotopy \cite{HatcherAT,FomenkoFuchs2016}.} Its classifying space has $\pi_i(B(K(A,1)))=\pi_{i-1}(K(A,1))$, therefore $B(K(A,1))\simeq K(A,2)$.\footnote{Notice therefore that also $K(A,2)$ can be taken as an Abelian group and $B(K(A,2))\simeq K(A,3)$. So $B^nA=K(A,n)$.} The principal bundles $BA \to BG \to BK$ are therefore classified by homotopy classes of maps $[BK,K(A,2)]$ and generically $[X,K(A,n)]\simeq H^n(X;A)$ \cite{HatcherAT}. This shows the $H^2(BK;A)$ classification of central extensions \eqref{general grp ext} for generic $K$.

\paragraph{Bockstein homomorphism.}
To any short exact sequence of Abelian groups \eqref{general grp ext}, there is an associate long exact sequence in cohomology \cite{HatcherAT} 
\begin{equation}
\label{LES Cohomology}
    ...\to H^n(X;A)\xrightarrow{\iota_*} H^n(X;G) \xrightarrow{\pi_*} H^n(X;K) \xrightarrow{\b_n} H^{n+1}(X;A)\to...
\end{equation}
for any topological space $X$. The connecting homomorphism is called Bockstein homomorphism. From the exactness of the sequence, $\ker(\b_n)=\Im(\pi_*)$: elements of $H^n(X;K)$ which can be lifted to elements of $H^n(X;G)$ are annihilated by the Bockstein, being therefore the obstruction to do such a lifting. Moreover, $\Im(\b_n)=\ker(\iota_*)$: $\iota_*:H^{n+1}(X;A)\to H^{n+1}(X;G)$ is not injective when $\b_n$ is not trivial. There is a simple way to construct such Bockstein map. Take a representative $x$ of an element $[x]\in H^n(X;K)$ and lift it to a cochain $\wt x$ in $C^n(X;G)$ such that $\pi(\wt x)=x$. Since $\d \pi(\wt x)=\pi (\d \wt x)=\d x=0$, we see that $\d \wt x \in \Im(\iota)$, therefore there is a cocyle $y\in Z^{n+1}(X;A)$ such that $\iota(y)=\d \wt x$. This is the Bockstein of $x$, i.e. $y = \b(x)$. $y$ is indeed zero when $x$ can be lifted to an element $\wt x \in Z^n(X;G)$, since $\d \wt x=0$.\footnote{More precisely, when $[y]=[\b(x)]=0$, so that $y=\d w$, it is possible to find a closed lift of $x$, by choosing $\wt x'=\wt x-\iota(w)$.} Notice that $\b(x)$ does not depend on the chosen representative: if we had chosen $x' = x+\d \l$, then $\wt x' = \wt x +\d \wt \l$ and $\d \wt x' = \d \wt x$, implying $\b(x+\d \l)=\b(x)$. However, $\b$ does depend on the choice of the lift. Consider a different lift $\wt x'$ of $x$, then $\pi(\wt x'-\wt x)=0$ and therefore $\wt x' = \wt x+ z$, with $z\in C^n(X;A)$.\footnote{More precisely, $\wt x'=\wt x+\iota_*(z)$, but we neglect $\iota$, according to our previous discussion. With the language used before, we are saying that $\wt x =s_*(x)$ for a choice of section $s: K\to G$, while $\wt x' =s'(x)$, for a different choice of section. Then $s'(x) = s(x)+\iota_* (z)$, with $z \in C^n(X;A)$. } As a consequence, $\b(x)$ depends on the choice of the lift for a coboundary term $\d z$. It is a well-defined homormophism in terms of cohomology classes, $[x]\mapsto [\b(x)]$.

As said, every central extension \eqref{general grp ext} is classified by an element $[\a]\in H^2(BK;A)$. Since the long exact sequence \eqref{LES Cohomology} is induced by \eqref{general grp ext}, one could expect that $\a$ specifies also \eqref{LES Cohomology}. Indeed, for example, if $[\a]=0$ and \eqref{general grp ext} splits, then $H^n(X;G)\simeq H^n(X;A\times K)\simeq H^n(X;A)\oplus H^n(X;K)$, therefore every element in $H^n(X;K)$ can be lifted to an element of $H^n(X;G)$ and all the Bockstein homomorphisms are trivial. More importantly, it is possible to define $\b_1$ in terms of $\a$ in a really straightforward manner. Given a map $\g: X \to BK$, it is possible to pullback $\a$ to $\g^*\a\in Z^2(X;A)$. When  $K$ is a finite Abelian group, $[X,BK]\simeq H^1(X;K)$ \cite{dijkgraafwitten}, with $BK=K(K,1)$,\footnote{We apologize for the incovenient notation, but $K(G,n)$ will be always specified by the group $G$ and the integer $n$, so no confusion should arise with the group $K$.} and it is natural to identify $\g^*\a = \a(\g)=\b_1(\g)$. So, one can see $\a$ as defining the Bockstein homomorphism $H^1(X;K)\to H^2(X;A)$. This connection between $\a$ and the Bockstein map is particularly useful when we consider the possible gauge fields for the extensions \eqref{general grp ext}. \\

\noindent \textit{Example.} Finite Abelian groups can always be written as a bunch of cyclic groups. We could therefore restrict the attention to sequences involving $\Z_n$ factors. A possible case is
\begin{equation}
\label{cyclic group ses}
    0\to\Z_n \xrightarrow{\cdot n} Z_{n^2}\xrightarrow{[]_n} \Z_n\to 0.
\end{equation}
The first map is multiplication by $n$, where $m\in \Z_n\mapsto nm \in \Z_{n^2}$, while $\pi = []_n$ is the modulo $n$ operation (clearly $\ker([]_n)=\Im(\cdot n)$). A nice property of \eqref{cyclic group ses} is that its Bockstein homomorphism $\b$ is a derivation \cite{HatcherAT}, i.e.
\begin{equation}
\label{bock derivation}
    \b(x \cup y) =  \b(x) \cup y+(-1)^p x\cup\b(y), \qquad [x]\in H^p(X;\Z_m),\;[y]\in H^q(X;\Z_m).
\end{equation} 
Indeed, starting with $x$, $y$ (a representative for each class), consider their lifts $\wt x$, $\wt y$ to cochains valued in $\Z_{n^2}$ (which can be thought of $x$ and $y$ themselves, where now they take values in the subset $0,...,n-1$ of $0,...,n^2-1$). We apply the map described above. Generally $\wt x$ and $\wt y$ are not closed, but $\d \wt x= n w$, $\d \wt y = n t$, with $w \in Z^{p+1}(X;\Z_n)$, $t \in Z^{q+1}(X;\Z_n)$. $w$ and $t$ are the obstructions to lift $x$ and $y$ to $\Z_{n^2}$-valued cocycles, i.e. $\b(x) = w = \d \wt x/n$ and $\b(y)=t= \d \wt y/n$. The lift of $x\cup y $ is $\wt x \cup \wt y$ and
\begin{equation}
    \d (\wt x \cup \wt y) = n (w \cup y + (-1)^p x\cup t) = n \b(x\cup y).
\end{equation}
This is exactly \eqref{bock derivation}.

\section{Abelian extensions of finite groups} 
\label{App GrpExt}

In this section all groups are finite and Abelian. We will use the additive notation for the group operation, when needed. 

\subsection{Physics of group extensions: gauge fields}
A gauge field\footnote{We denote the gauge field with the same letter of the group.} $G$ for the extension \eqref{general grp ext} can be written with a pair $(A,K)$ with the constraint 
\begin{equation}
\label{gauge field grp ext}
    \d A = K^*\a = \a(K),\qquad A\in C^1(X;A),\;K\in Z^1(X;K).
\end{equation}
There are various ways to see this. The most straightforward is to consider the dual network of symmetry defects and the product rule \eqref{grp ext product}. This shows that when two $K$ symmetry defects fuse, an $A$-symmetry defect $\a(k_1,k_2)$ appears at their junction; in terms of the dual background gauge fields, this is \eqref{gauge field grp ext}, which means that $K$ is a source for $A$. Therefore, the gauge field $K$ is still closed, $[K]\in H^1(X;K) \simeq [X,BK]$, while $A \in C^1(X;A)$ in general, since $A$-defects could have boundaries when $K\neq 0$. 

From a more geometrical point of view, one could start with a gauge field $[K]\in H^1(X;K)$ and asks if it can be lifted to a gauge field for the whole $G$. Concretely, given a gauge field $K$ that satisfies the cocycle condition on triple intersections of patches, $k_{ij}k_{jk}k_{ki}=1$ (with $k_{ij}$ the transition function on $U_i \cap U_j$), we would like to construct a gauge field $G$, such that $\pi(g_{ij}) = k_{ij}$, that still satisfies the cocycle condition. Generally, the lifted gauge field satisfies instead $g_{ij}g_{jk}g_{ki}=\b_{ijk}$, with $[\b]\in H^2(X;A)$ in the kernel of $\pi$ that is the obstruction to do such lifting. This object is indeed the Bockstein of $K$: when it is trivial in cohomology, $\b_{ijk}=(\d a)_{ijk}$, it is possible to consistently lift $K$ to a gauge field for the whole group $G$, by choosing $g_{ij}'=a_{ij}^{-1}g_{ij}$. This is exactly the content of \eqref{gauge field grp ext}, where the object that trivializes $\b_1(K)$ is the gauge field for $A$. Notice that when the lift is possible, there are $|H^1(X;A)|$ inequivalent lifts, given by $a_{ij}^{-1}g_{ij} \to s_{ij}a_{ij}^{-1}g_{ij}$, with $[s]\in H^1(X;A)$. This shows that the datum of $A$ in \eqref{gauge field grp ext} is actually needed to specify $G$ (not just $K$ and $\b(K)$).

This latter point of view can be reformulated directly in terms of gauge fields, which allows for a generalization to higher-form symmetries next. We know from the start that $[G]\in H^1(X;G)$ and, given the extension \eqref{general grp ext}, it can be written as
\begin{equation}
\label{gauge field for G}
    G= \Tilde{K}-\iota(A), \qquad A\in C^1(X;A),\;\pi(\Tilde{K})=K\in C^1(X;K).
\end{equation}
From $\d G=0$ and applying $\pi$, it follows that $\d K=0$, so $K$ is a cocycle. We also find that $\d \Tilde{K}$ is in $\ker(\pi)=\Im(\iota)$, therefore there should exist an element $x$ in $Z^2(X;A)$ such that $\iota(x) = \d \Tilde{K}$: the assignment of $x$ to $K$ is the role of the Bockstein map and then $\d A =x$. This is \eqref{gauge field grp ext}. We are saying, again, that, starting with $[K]\in H^1(X;K)$, it is possible to consistently construct $[G]\in H^1(X;G)$ according to \eqref{gauge field for G}, provided that $A$ trivializes the Bockstein of $K$ (the inequivalent lifts mentioned above are then $G\to G'= \wt K - \iota(A+s)$, with $[s]\in H^1(X;A)$).

We finally consider the gauge transformations for the pair $(A,K)$, which should be compatible with the constraint \eqref{gauge field grp ext}. To derive them, it is more convenient to start from the whole $G$ field. The standard gauge transformation for $G\in Z^1(X;G)$ is $G\to G+\d\L$, $\L \in C^0(X;G)$. This descends to the standard gauge transformations for $A$ and $K$ in \eqref{gauge field for G}, by writing $\L =\wt\l-\iota(\g)$, with $\pi(\wt \l)=\l$, $\l \in C^0(X;K)$, $\g \in C^0(X;A)$. It gives $A\to A+\d \g$ and $K\to K+\d\l$, which do not modify \eqref{gauge field grp ext}.\footnote{In terms of transition functions, $\l$ corresponds to a gauge transformation for $k_{ij}\to \l_i^{-1}k_{ij}\l_j$, which yields a gauge transformation for the lift $g_{ij}\to \wt \l_i^{-1}g_{ij}\wt\l_j$. There are also gauge transformations for $g_{ij}$ given by $\iota(\g)$ which are not seen by $k_{ij}$ and that modify $a_{ij}$.} However, the splitting of $G$ in terms of $(A,K)$ as in \eqref{gauge field for G} is not unique and there can be other choices, given by\footnote{Equivalently, there are various way to split the transitions functions $g_{ij}=a_{ij}^{-1}\wt k_{ij}=a_{ij}'^{-1}\wt k'_{ij}$, such that $\pi(g_{ij})=k_{ij}$. If $\wt k_{ij}$ gives the cocycle condition with $(\d a)_{ijk}$, $\wt k_{ij}'= \w_{ij} k_{ij}$ is another lift with cocycle condition giving $(\d a+\d \w)_{ijk}=(\d a')_{ijk}$.}
\begin{equation}
\label{different splitting G}
    G= \wt K-\iota(A)=\wt K'-\iota(A') \quad \Rightarrow \quad \wt K'= \wt K+\iota(\w), \; A'=A+\w, \; \w\in C^1(X;A).
\end{equation}
They correspond to different choices for the lift $K\to \wt K$. From the point of view of the whole symmetry $G$, these are extra gauge transformations that arise when we represent $G$ as a pair $(A,K)$. The transformations \eqref{different splitting G} still leave invariant the constraint \eqref{gauge field grp ext}, since $\a(K)\to \a(K)+\d \w$ which matches the variation of $A$. Notice that these are all the gauge transformations that leave invariant the condition \eqref{gauge field grp ext}, the last one due to the fact the $\b(K)=\a(K)$ is not uniquely defined by $K$, but depends on a choice of its lift. One could also think of this as a large gauge transformation for $K$: lifting $K \to \wt K$, we have the following gauge transformation $\wt K \to \wt K +\d \wt \l+\iota(\w)$.

All in all, the gauge transformations for $(A,K)$ compatible with \eqref{gauge field grp ext} are therefore
\begin{equation}
\label{gauge transf grp ext}
    K \to K^\l= K +\d \l, \quad \wt K\to \wt K+\d\wt \l+\iota(\w)\quad A \to A+\d \g + \omega, \quad \a(K)\to \a(K)+\d \w.
\end{equation}
Notice that $\w$ is the non-trivial content of the extension \eqref{general grp ext}, while the other gauge transformations $A\to A+\d \g$ and $K\to K+\d \l$ are there also for a standard product symmetry $A\times K$. For the following, it will be convenient to include the large gauge transformation $\w$ implicitly in the compact expression $K\to K^\l$. This simplifies the notation and allows for a more direct generalization to the higher group case considered next. From this point of view, $\a(K^\l)-\a(K)=\d \w$, which shows the $\w$ is a first descend of $\a$ ($\delta \a =\d \w$).

\subsection{Higher form symmetries and higher groups} 

The precedent discussion can be generalized to group extensions of higher-form symmetries. This is actually required to consistently treat gauging of group extensions, since gauging an ordinary zero-form symmetry produces a dual higher-form symmetry in general.  

The main physical content of a group extension can be encoded in terms of  the background gauge fields as in \eqref{gauge field grp ext}. We therefore say, more generally, that a symmetry $K^{(p)}$ is extended by another symmetry $A^{(q)}$ if their gauge fields satisfy\footnote{See also Appendix A of \cite{AntoSantaniello2025}.}
\begin{equation}
\label{gauge field grp ext higher form}
    \d A = \a(K), \qquad K\in Z^{p+1}(X;K), \qquad A\in C^{q+1}(X;A),
\end{equation}
with a cohomological operation $\a$ of type $(p+1,q+2,K,A)$, i.e. a natural\footnote{For every continuous map $f: X\to Y$, $f^*\a_Y=\a_Xf^*$.} map
\begin{equation}
\label{alpha as cohomol op}
    \a : H^{p+1}(-;K)\to H^{q+2}(-;A).
\end{equation}
Cohomological operations of this type are in bijective correspondence with the cohomology group $H^{q+2}(K(K,p+1);A)$ \cite{HatcherAT,FomenkoFuchs2016}.\footnote{This follows easily from the isomorphism $H^n(X;A)\simeq [X,K(A,n)]$ \cite{HatcherAT,FomenkoFuchs2016}.} Notice here that $K(G,p+1)=B^{p+1}G$ is the classifying space for a $p$-form symmetry $G$. In the case $p=q=0$ we recover the standard central extension of before, classified by $[\a]\in H^2(BK;A)$ \eqref{general grp ext}. Strictly speaking, when $q$ is bigger then $p$, the resulting structure is not quite a group, but a higher-group \cite{KapustinThorngren2013HigherGroup}. For example, when $p=0$ and $q=1$, we have a 2-group classified by $[\a] \in H^3(BK;A)$, which is called Postnikov class \cite{CordovaDumitrescu2groups,CordovaBeniniHsin2group}.

Cohomological operations like \eqref{alpha as cohomol op} are not necessarily group homomorphisms, which is a main difference with the group extension case (where \eqref{alpha as cohomol op} is the Bockstein for \eqref{LES Cohomology}, thus a homomorphism). The pair $(A,K)$ has still gauge transformations analogous to \eqref{gauge transf grp ext}, where $\w$ is a first descend of $\a$, which encodes the fact that we are working with representative cocycles and not cohomology classes. However, in principle here it could depends on $K$ and $\l$ themselves, so $\w=\w(K,\l)$ and $\a(K+\d\l)=\a(K)+\d \w(K,\l)$ even for small gauge transformations.

The relation \eqref{gauge field grp ext higher form} can be understood in terms of the dual symmetry defects, as before. Equation \eqref{gauge field grp ext higher form} tells that when a suitable number of $K$-defects meet, at their junction an $A$-symmetry defect starts. $K$ is thus a source for $A$. By dimensional reasons,  $q-p+2$  $(d-p-1)$-dimensional defects of $K$
should meet to make a source for a $(d-q-1)$-dimensional defect for $A$. We deduce that in general $q \geq p$ to make a sensible extension and indeed $H^{m}(K(K,n);A)=0$ when $m<n$. There is still the possibility of a $p$-form symmetry extended by a $(p-1)$-form symmetry, where every defect of $K^{(p)}$ by itself is a source for $A^{(p-1)}$. In such a case, since $H^{n}(K(K,n);A)\simeq \hom(K,A)$, the cohomology operation \eqref{alpha as cohomol op} is just a group homomorphism $\phi: K \to A$ such that $\d A = \phi(K)$. Notice that the physical point of view in terms of symmetry defects is somewhat trivial when $K$ is a zero-form symmetry, since $H^n(K(K,1);A) \simeq H^n_{grp}(K;A)$. In the general case is not obvious mathematically.\footnote{To the knowledge of the author.}

The case $q=p$ can be thought of as a really direct generalization of the group extension \eqref{general grp ext}, where the higher-form nature of the objects do not play a particular role. In such a  case we have really a group extension,
\begin{equation}
\label{grp ext higher-form}
       0 \to A^{(p)} \xrightarrow{\iota} G^{(p)} \xrightarrow{\pi} K^{(p)} \to 0, \qquad [\a]\in H^{p+2}(B^{p+1}K;A).
\end{equation}
The object that classifies this extension is a natural generalization of the zero-form symmetry case, which indeed gives a map $H^{p+1}(X;K)\to H^{p+2}(X;A)$ given by $K \to\a(K)= K^*\a$, that is interpreted as the Bockstein $\b_p(K)$. Notice also that $H^{m+1}(K(K,m);A)$, for $m \geq 2$, classifies the Abelian extensions of $K$ by $A$ \cite{EilenbergMacLaneOriginal}, so there should be an isomorphism $H^{m+1}(K(K,m);A) \simeq H^2_{sym}(K(K,1);A)$. This shows that the information of \eqref{grp ext higher-form} is really equivalent to that of \eqref{general grp ext}. Applying the very same argument around \eqref{gauge field for G}, but to $p+1$ cohomology groups, with $[G]\in H^{p+1}(X;G)\simeq [X,K(G,p+1)]$, one gets a pair $(A,K)$ with the constraint $\d A = \b_p(K)$, which is \eqref{gauge field grp ext higher form}.

\subsection{Pontryagin dual sequence}
In general, gauging a finite Abelian symmetry $G$ produces a dual $(d-p-2)$-form symmetry $\wh G = \hom(G,\U(1))$ \cite{tachikawaGaugeFiniteGroups,bhardwajTachikawa2017}, which is the Pontryagin dual group of $G$, i.e. the group of characters of $G$ or, equivalently, the group of its irreducible representations. For finite Abelian groups, $\wh G \simeq G$ (albeit not canonically). It is thus interesting to look at the Pontraygin dual of the extension \eqref{general grp ext}.

Even if $\wh G \simeq G$, the extension presentation \eqref{general grp ext} of $G$ is inverted under the Pontryagin duality operation, it becomes 
\begin{equation}
\label{pontryagin dual ext}
     0 \to \wh K \xrightarrow{\wh \pi} \wh G \xrightarrow{\wh\iota} \wh A \to 0, \qquad [\wh\a] \in H^2(B\wh A;\wh K).
\end{equation}
To see this, consider a character $\chi \in \wh G$, i.e. a map $\chi: G\to \U(1)$. According to \eqref{general grp ext}, this naturally gives a character for $A$ by simply restricting $\chi$ to $\iota(A)\simeq A$ (or, in other words, $\chi_A = \chi \circ \iota$ is a well-defined character for $A$, given  that $\iota$ is injective). This gives a surjective map $\wh \iota: \wh G \to \wh A$. The kernel of this map are the characters of $G$ that do not restrict to characters for $A$, so they are characters for $G/A \simeq K$. Therefore, $\ker(\wh \iota) = \wh K$. If we define $\wh \pi$ as an injective inclusion $\wh K \hookrightarrow \wh G$ we finally get the dual extension \eqref{pontryagin dual ext}. By the general theory of group extensions, \eqref{pontryagin dual ext} is classified by a symmetric class $[\wh\a] \in H^2(K(\wh A,1);\wh K)$.

The dual sequence \eqref{pontryagin dual ext} comes from \eqref{general grp ext}, so one could argue that $\wh \alpha$ is defined by $\a$.\footnote{See the question and the nice answer at \url{https://mathoverflow.net/questions/450492/pontryagin-dual-of-a-group-cohomology-class}.} However, having in mind a physical application, notice the dual sequence \eqref{pontryagin dual ext} that arises after gauging a zero-form symmetry \eqref{general grp ext} in $d$-dimensions is really a sequence for $(d-2)$-form symmetries like \eqref{grp ext higher-form}, with $[\wh \alpha]\in H^d(K(\wh A,d-1);\wh K)$. As noted before, there should be an isomorphism $H^{m+1}(K(\wh A,m);\wh K) \simeq H^2_{sym}(K(\wh A,1);\wh K)$. If we take $m = d-1$, we see that specifying an element in $H^2_{sym}(K(\wh A,1);\wh K)$ suffices, together with an explicit isomorphism between the two groups. For us, it is however more convenient to define directly an object in $H^d(K(\wh A,d-1);\wh K)$. Moreover, for  reasons that will become clear in the next section, we actually want a relation between $\a$ in $\wh \a$ in $(d+1)$-dimensions. So, our aim is to understand how a cohomological operation $\a: H^1(X;K)\to H^2(X;A)$ gives rise to a dual operation $\wh \a: H^{d-1}(X;\wh A) \to H^{d}(X;\wh K)$ on every $(d+1)$-dimensional manifold $X$.

Take $X$ $(d+1)$-dimensional and $[k] \in H^1(X;K)$. Using $\a$, one can construct a pairing $[\a(k)]\cup [\wh a]$ with $\wh a \in H^{d-1}(X;\wh A)$.\footnote{The cup product is constructed with the natural pairing $A \times \wh A \to \R/\Z$ \cite{Steenrod1947}.} Because $\iota$ is injective at the cochain level, this can be written as $\iota(\a(k)) \cup \widetilde{\wh a}$ for a lift $\widetilde{\wh a}$ of $\wh a$ to a $\wh G$ valued cochain for the dual sequence.\footnote{By construction of the dual sequence \eqref{pontryagin dual ext}, a character $\wh k$ (as an element of $\wh G$) evaluates to zero on an element of $A$ (as an element of $G$). Therefore $\iota(\a(k)) \cup \widetilde{\wh a}$ is independent from the choice of the lift.} Since $\alpha=\beta$, by evaluating on the fundamental class gives 
\begin{equation}
\label{f for Znm}
\begin{split}
    \int_X \b(k) \cup \wh a =\int_X \iota(\b(k)) \cup \widetilde{\wh a} =\int_X \d \wt k\cup \widetilde{\wh a} = \int_X \wt k \cup \d \widetilde{\wh a} +\d( \wt k\cup \widetilde{\wh a} )=\\= \int_X \wt k \cup \wh \iota(\wh \b(\wh a))+\d (\wt k\cup \widetilde{\wh a}) = \int_X  k \cup \wh \b(\wh a)+\d ( \wt k\cup \widetilde{\wh a}),
\end{split}
\end{equation}
where $\wh \b: H^{d-1}(X;\wh A) \to H^{d}(X;\wh K)$ is the  Bockstein homomorphism for \eqref{pontryagin dual ext}. This gives the natural cohomological operation we were looking for $\wh\a \in H^d(K(\wh A,d-1);\wh K)$, which is just a consequence of integration by parts. We thus obtained 
\begin{equation}
\label{hat alpha from alpha}
 \a(k) \cup \wh a =k \cup \wh \a(\wh a)+\d ( \wt k\cup \widetilde{\wh a}),
\end{equation}
where the exact piece is a correction term that vanishes in cohomology.

Notice that the very same argument can be made starting with a sequence of $p$-form symmetries in \eqref{general grp ext}. Then the dual sequence \eqref{pontryagin dual ext} involves $(d-p-2)$-form symmetries and \eqref{hat alpha from alpha} defines $\wh \a: H^{d-p-1}(X;\wh A) \to H^{d-p}(X;\wh K)$ in terms of $\a: H^{p+1}(X;K)\to H^{p+2}(X;A)$ on a $(d+1)$-dimensional $X$. This is instead not generically true for higher-groups, where $\a$ in \eqref{hat alpha from alpha} is not a Bockstein homomorphism.

\subsection{Gauging in two steps}

In this section we will see the effect of gauging a non-anomalous Abelian group extension of the form \eqref{general grp ext}. The effect of gauging $A$ is already considered in detail in \cite{tachikawaGaugeFiniteGroups}. The fact that gauging sub-sequentially $A$ and then $K$ is equivalent to gauging the whole $G$ is shown in \cite{ThomasNonInvSymRepsII}: there, the authors restrict to two-dimensions, but they do it in full generality by considering also non-Abelian groups, which requires the language of fusion categories. Here we restrict to the Abelian case and we will formulate the argument in terms of partition functions, in general dimension $d$. We will see explicitly how the dual sequence \eqref{pontryagin dual ext} arises, with the exchanged roles for $A$ and $K$.

Consider a theory $\T$ with $G$ global symmetry, which is not anomalous,
\begin{equation}
    Z[A,K] = Z[A+\d \g + \w,K+\d\l].
\end{equation}
Now we gauge the symmetry $A$ to obtain $\T/A$. The gauged theory has a dual $(d-2)$-form symmetry $\wh A$ with partition function
\begin{equation}
    Z/A[\wh A, K]= \sum_{a|\d a= \a(K)} Z[a,K]e^{2\pi i \int a \cup \wh A}.
\end{equation}
Gauge invariance for $a$ requires $\d \wh A = 0$. The symmetry in the gauged theory is therefore $\wh A^{(d-2)}\times K$, but they have a mixed anomaly: under the background gauge transformation for $K$ and $\wh A \to \wh A + \d \wh \g$, $Z/A$ changes by a phase
\begin{equation}
    e^{2\pi i \int_X \a(K) \cup \wh \g+ \w \cup \wh A + \d \w \cup\wh \g}.
\end{equation}
This phase comes from the variation of the $d+1$ dimensional SPT phase (anomaly theory)
\begin{equation}
\label{group extension anomaly app}
    2\pi i\int_Y \a(K) \cup \wh{A}, \qquad \p Y = X.
\end{equation}
This anomaly comes from the condition \eqref{gauge field grp ext higher form} (therefore applies also for higher groups).\footnote{At the level of symmetry defects, notice that the dual symmetry $\wh A^{(d-2)}$ is generated by the Wilson lines of $a$ after gauging \cite{bhardwajTachikawa2017}. However, when \eqref{gauge field grp ext} holds, these Wilson lines are not topological when $K \neq 0$, but they change with a phase given by $\a(K)$ after a continuous deformation. This is another point of view for the anomaly \eqref{group extension anomaly app}.} Rather then preventing to gauge $\wh A$ when $K \neq 0$, it is actually the piece of information that remembers (or says, if one starts from $\T/A$) that the global symmetry of $\T$ is the non-trivial extension \eqref{general grp ext}. Indeed, for a given $K$ with $\a (K) =\d A'$, it can be trivialized by the addition of a local counterterm $A'\cup\wh A$ in $Z/A$. This allows to gauge back $\wh A$ and $\T/A/\wh A \simeq \T$. In $\T$, $A'$ becomes the background gauge field for $A$. 

We now gauge also $K$ to obtain $\T/A/K$ and we want to show that $\T/A/K \simeq \T /G$. The final theory should have a dual symmetry $\wh G^{(d-2)}$, given by \eqref{pontryagin dual ext}, but applied to $(d-2)$-form symmetries. As said, this is classified by $[\wh \a] \in H^d(B^{d-1}\wh A,\wh K)$ \cite{tachikawaGaugeFiniteGroups}, which defines a map $H^{d-1}(X,\wh A) \simeq [X, B^{d-1} \wh A]\to H^d(X,\wh K)$ such that $\d \wh K =\wh \a (\wh A)$. Recall also that $\wh G$ is non-anomalous and by gauging it one can recover $\T$. Let us see how this works.

A crucial fact is that the anomaly \eqref{group extension anomaly app} is symmetric in $K$ and $\wh A$, because of the relation \eqref{hat alpha from alpha}, i.e.
\begin{equation} 
\label{anom grp ext = anom dual grp ext}
    \int_Y \a(K)\cup \wh A= \int_Y K \cup \wh\a(\wh A)+ \int_X \wt K\cup \widetilde{\wh A}, \qquad \p Y=X.
\end{equation}
Namely, the two anomaly actions define the same anomaly, up to a local counterterm on $X$ given by $f(K,\wh A) = \wt K\cup \widetilde{\wh A}$. Notice that this term does depend on the lifts and under $\wt K\to \wt K +\d\wt \l+\w$ it changes as\footnote{Namely, $\d \wt \l \cup \widetilde{\wh A}$ requires a lift of $\wh A$, while $\w \cup \widetilde{\wh A}=\w \cup \wh A$ does not depend on the choice of the lift.}
\begin{equation}
\label{property of f for alpha}
    (\wt K +\d\wt \l+\w)\cup \widetilde{\wh A}-\wt K\cup \widetilde{\wh A} = \d \wt \l \cup \widetilde{\wh A}+ \w\cup \wh A.
\end{equation}

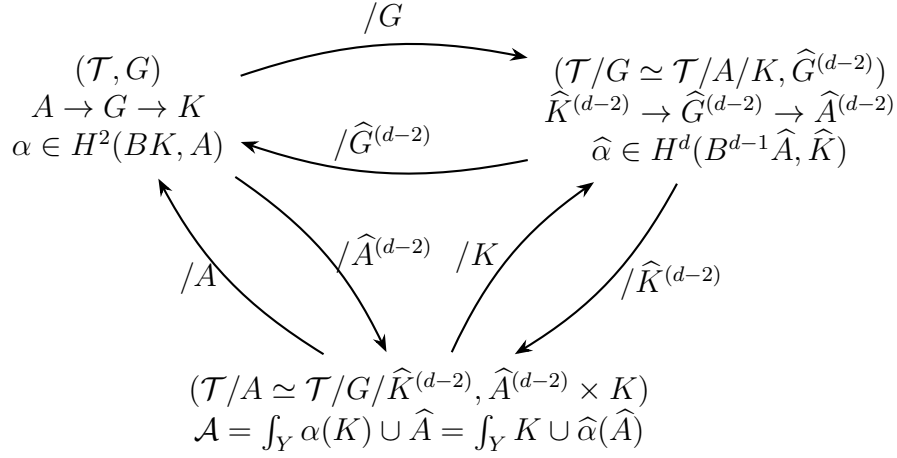
\begin{figure}
\centering
\begin{tikzpicture}[
    every node/.style={align=center},
    arrow/.style={-{Stealth}, thick},
    shiftarr/.style={shorten >=2pt, shorten <=2pt}
]

% Triangle points (pointing downward)
\node (A) at (0, 0) {$(\T/A \simeq \T/G/\wh K^{(d-2)},\wh A^{(d-2)} \times K)$\\$\A= \int_Y \a(K)\cup \wh A= \int_Y K \cup \wh\a(\wh A)$};
\node (B) at (-4, 4) {$(\T,G)$\\$A\to G \to K$ \\ $\a \in H^2(BK,A)$};
\node (C) at (4, 4) {$(\T/G\simeq \T/A/K,\wh G^{(d-2)})$\\$\wh K^{(d-2)}\to \wh G^{(d-2)} \to \wh A^{(d-2)}$ \\ $\wh \a \in H^d(B^{d-1}\wh A,\wh K)$};

% Offset for parallel arrows
\def\offset{0.3}

% Arrows between B <-> C
\draw[arrow, shiftarr] (B) to[bend left=15] node[above] {$/G$} (C);
\draw[arrow, shiftarr] (C) to[bend left=15] node[above] {$/\wh G^{(d-2)}$} (B);

% Arrows between C <-> A
\draw[arrow, shiftarr] (C) to[bend left=15] node[right] {$/\wh K^{(d-2)}$} (A);
\draw[arrow, shiftarr] (A) to[bend left=15] node[left] {$/K$} (C);

% Arrows between A <-> B
\draw[arrow, shiftarr] (A) to[bend left=15] node[left] {$/A$} (B);
\draw[arrow, shiftarr] (B) to[bend left=15] node[right] {$/\wh A^{(d-2)}$} (A);

\end{tikzpicture}
    \caption{Gauging an Abelian extension $A\to G\to K$ in one or two steps.}
    \label{diagram}
\end{figure}

To define the partition function of $\T/A/K$, we first subtract the counterterm $f$ (basically shifting the anomaly to $K\cup \wh \a(\wh A)$) and then we sum over $K$, obtaining
\begin{equation}
\label{gauged full grp ext}
        Z/A/K[\wh A, \wh K]= \sum_{k,a|\d a= \a(k)} Z[a,k]e^{2\pi i \int a \cup \wh A+k \cup \wh K-\wt k\cup \widetilde{\wh A}}.
\end{equation}
The counterterm $f$ is needed to ensure gauge invariance under \eqref{gauge transf grp ext}. Indeed, after a gauge variation \eqref{gauge transf grp ext}, we get
\begin{equation}
    \d \wh A =0,\qquad \int_X \omega\cup\wh A+\d\l\cup \wh K-(\d \wt \l \cup \widetilde{\wh A}+ \w\cup \wh A )=0.
\end{equation}
We have already found the first condition after gauging $A$. The second one gives
\begin{equation}
       \int_Y \d\l\cup \d\wh K -\d\wt \l \cup \d \widetilde{\wh A}=  \int_Y \d \l \cup\d\wh K-\d\wt \l\cup \wh\pi(\wh\b(\wh A))=\int_Y \d \l \cup(\d\wh K-\wh\b(\wh A))=0,
\end{equation}
where the last equality follows from the same reasoning explained above \eqref{f for Znm}. So, by suitable choosing the counterterm $f(K,\wh A)$ and identifying again $\wh \b=\wh \a$, gauge  invariance of $T/A/K$ requires
\begin{equation}
    \d \wh A= 0, \qquad \d \wh K =\wh\a(\wh A).
\end{equation}
This is indeed the statement that $T/A/K$ has a symmetry $\wh G^{(d-2)}$ given by the extension \eqref{pontryagin dual ext}. Gauging $\wh K^{(d-2)}$ gives back the theory $\T/A \simeq T/A/K/\wh K \simeq \T/G/\wh K$. This is shown in the diagram in Figure \ref{diagram}.\footnote{Notice that in the middle step, we obtain that $\T/G/\wh K \simeq \T/A$ with a tuning of counterterms. This is expected, since, starting from $\T/G$, one obtains naturally an anomaly in terms of $\wh \a$, which differs from \eqref{group extension anomaly app} by precisely such terms.}

We could also check that $\wh G^{(d-2)}$ is not anomalous as a whole, so that $Z/A/K$ must be invariant under
\begin{equation}
\label{gauge trans dual grp ext}
   \wh A \to A^\g= \wh A +\d \wh \g, \qquad \wh K \to \wh K +\d \wh \l + \wh \omega, \qquad \d \wh \omega= \wh \a(\wh A^\g)-\wh \a(\wh A)
\end{equation}
(the analogous of \eqref{gauge transf grp ext} for \eqref{pontryagin dual ext}, compatible with $\d \wh K =\wh \a(\wh A)$). This is the case when $\d k=0$ and $\d a = \a(k)$, which are indeed our starting hypothesis. 

Note that everything applies in the same way also if we start with an extension of $p$-form symmetries \eqref{grp ext higher-form}. We still get the very same diagram above, with the only difference that the dual symmetries are $(d-p-2)$-form symmetries and $\a$ and $\wh \a$ are modified accordingly to the degree. This is instead not generically true for higher-groups, where the crucial difference is that the anomaly \eqref{group extension anomaly app} is not symmetric as in \eqref{anom grp ext = anom dual grp ext}. Generically, the result of gauging a higher-group is a non-invertible symmetry (see \cite{BasonCuiRuggeri2026} for an explicit example).

\subsection{Gauging in one step}

In the previous section we showed how gauging subsequentially $A$ and then $K$ in \eqref{general grp ext} yields a theory with symmetry group $\wh G$ \eqref{pontryagin dual ext}. This was done by an explicit step by step computation, showing that the final symmetry data are indeed the ones of \eqref{pontryagin dual ext}. We now gauging directly $G$ knowing that the dual symmetry should be \eqref{pontryagin dual ext}.

The partition function of $\T/G$ is 
\begin{equation}
    Z/G[\wh G]= \sum_{g} Z[g]e^{2\pi i \int g \cup \wh G}.
\end{equation}
Using the decomposition \eqref{gauge field for G}, namely
\begin{equation}
    G = \wt K -A,\qquad \wh G = \wh K - \widetilde{\wh A},
\end{equation}
this can be written as 
\begin{equation}
    Z/G[\wh G = (\wh A,\wh K)]= \sum_{g=(k,a)} Z[g=(k,a)]e^{2\pi i \int a \cup  \widetilde{\wh A}+ \wt k \cup \wh K - \wt k\cup \widetilde{\wh A}- a \cup \wh K}.
\end{equation}
By construction of the dual sequence \eqref{pontryagin dual ext}, a character $\wh K$ (as an element of $\wh G$) evaluates to zero on an element of $A$ (as an element of $G$). Therefore, the term $a\cup \wh K$ above is trivial, while the first two terms do not depend on the lifts (sending $\widetilde{\wh A}\to \widetilde{\wh A}+\wh K$ and $\wt k \to \wt k +a$ leave them invariant). The final partition is thus 
\begin{equation}
    Z/G[\wh G = (\wh A,\wh K)]= \sum_{g=(k,a)} Z[g=(k,a)]e^{2\pi i \int a \cup  \wh A+ k \cup \wh K - \wt k\cup \widetilde{\wh A}}.
\end{equation}
This is exactly \eqref{gauged full grp ext} with the counterterm $f(K,\wh A)= \wt K \cup \widetilde{\wh A}$. This is a confirmation of the diagram in Figure \ref{diagram}.

\section{Cyclic extensions of $\U(1)$ and their gauging}
\label{U(1) extensions sec}

\subsection{Extension}

Consider a theory $\T$ with a $\U(1)$ symmetry. We want to gauge first a subgroup $\Z_q\subset \U(1)$ and then the remaining group $\U(1)/\Z_q \simeq \wt \U(1)$ (albeit $\wt \U(1)\simeq \U(1)$ topologically). This can be understood by decomposing $\U(1)$ as the extension
\begin{equation}
\label{U(1) grp ext}
    \Z_q \xhookrightarrow{\iota}  \U(1) \xrightarrow{\pi} \wt \U(1) \simeq \U(1)/\Z_q, \qquad [\a] \in H^2(B\wt\U(1);\Z_q) \simeq \Z_q,
\end{equation} 
where $B\U(1)  \simeq K(\Z,2)$ is the classifying space for $\U(1)$, $K(\Z,2)$ being a Eilenberg-MacLane space, and $H^2(K(\Z,2);\Z_q)\simeq \hom(\Z,\Z_q)\simeq\Z_q$ \cite{HatcherAT,FomenkoFuchs2016}.\footnote{Fix $\pi:  u \in \U(1) \mapsto u^q=\wt u \in\wt\U(1)$, the $q$ inequivalent extensions \eqref{U(1) grp ext} are given by $\iota_k: n \in \Z_q \mapsto e^{2\pi kn/q}$ (i.e. each element of $\Z_q$ is sent to a $q$--th rooth of unity and indeed $\Im(\iota_k)=\ker(\pi)$), with $k =0,1,...,q-1 \in \Z_q$, $k=0$ being the split extension $\U(1) \times \Z_q$.} Physically, the difference between $\U(1)$ and $\wt\U(1) \simeq \U(1)/\Z_q$ is a relative factor of $q$ in their fundamental charges: since $\wt\U(1)$ is obtained after gauging $\Z_q\in \U(1)$, if $e=1$ is the fundamental charge of $\U(1)$ then the minimal charge probed by $\wt\U(1)$ is $q e =q$.

The extension \eqref{U(1) grp ext}, in terms of bundles, implies that $q[c_1(A)] = [c_1(\wt A)]$,\footnote{The homomorphism $\pi$ in \eqref{U(1) grp ext} induces a map at the level of transition functions $u_{ij}: U_i\cap U_j \to  U(1) \mapsto  u^q_{ij}=\wt u_{ij}: U_i \cap U_j \to \wt \U(1)$. The cocyle condition $u_{ij} u_{jk} u_{jk}=1$, from which one may extract $c_1(A)$, becomes $(u_{ij} u_{jk} u_{jk})^q=1$, yielding $qc_1(A)= c_1(\wt A)$.} where $A$ is the background for $\U(1)$ and $\wt A$ for $\wt U(1)$. This says that, given $S \in C^1(X;\Z_q)$ a background for $\Z_q$, we have
\begin{equation}
\label{U(1) ext gauge fields}
    \d S = c_1(\wt A)|_q,
\end{equation}
analogous of \eqref{gauge field grp ext}.\footnote{Notice that \eqref{U(1) ext gauge fields} is the same of \eqref{gauge field grp ext} since, if $\wt A\in [X,B\wt\U(1)]$ is the gauge field for $\wt\U(1)$, then the class $\wt A^*[\a] \in H^2(X;\Z_q)$ is really $\wt A^*[\a]=[c_1(\wt A)]_q$, given that any $\U(1)$ bundle is classified uniquely by its first Chern class $c_1$.} This gives a decomposition of $A$ into the pair $(S,\wt A)$, which can be lifted to 
\begin{equation} 
\label{U(1) ext gauge fields lift}
    q c_1(A) = c_1(\wt A)-\d \wt S,
\end{equation}
for an integer lift $\wt S$ of $S$. The gauge transformations compatible with \eqref{U(1) ext gauge fields} are
\begin{equation}
\label{extra gauge transf for U(1) ext}
    c_1(\wt A) \to c_1(\wt A) +\d n, \qquad S\to S+\d \g+ n|_q, \qquad n \in C^1(X;\Z).
\end{equation}
Note that these are all conditions on the global part of the $\U(1)$ data, since the extension by $\Z_q$ just regards the topological part. Locally, $q A = \wt A$. The global information can also be encoded in the holonomies of $A$, written as
\begin{equation}
\label{wilson line A Atilde S}
    e^{i\oint A} = e^{\frac{i}{q}\oint \wt A}e^{-\frac{2\pi i}{q}\oint S}, 
\end{equation}
which are invariant under \eqref{extra gauge transf for U(1) ext}. Improperly quantized Wilson lines of $\wt A$ can be dressed by $S$ satisfying $\d S= \d \wt A/2\pi|_q$ to give the Wilson lines for $A$.
 
The result \eqref{group extension anomaly app} is completely general  and it always applies for \eqref{general grp ext} for a generic group $K$ when $A$ is a finite Abelian group \cite{tachikawaGaugeFiniteGroups}; so it applies also to \eqref{U(1) grp ext}. Indeed, the condition \eqref{U(1) ext gauge fields} implies that, after gauging $\Z_q$, the dual $\wh \Z_q^{(d-2)}$ symmetry with gauge field $C\in Z^{d-1}(X;\Z_q)$ and the remaining $\wt\U(1)$ factor have the mixed anomaly 
\begin{equation}
\label{U(1) grp ext anom}
        \frac{2\pi i}{q} \int_Y c_1(\wt A)|_q \cup C =\frac{2\pi i}{q} \int_Y c_1(\wt A) \cup \wt C,
\end{equation}
where the second expression is a lift to integer-valued chains, $\wt C \in C^{d-1}(X;\Z)$ (notice that in this Section we always pick an isomorphism for the dual symmetry, such that $\wh \Z_q \simeq \Z_q$). This can be understood from \eqref{wilson line A Atilde S}: the Wilson lines for $S$ are not strictly topological when $\wt A \neq 0$, which is the origin of the mixed anomaly \eqref{U(1) grp ext anom} from a symmetry defect perspective.

As for the finite groups, we want to understand what happens if we gauge also $\wt \U(1)$, which should lead to $\T/\Z_q/\wt\U(1)\simeq\T/\U(1)$. The final theory should be characterized generically by just the magnetic $\U(1)^{(d-3)}_m$ symmetry \cite{gaiotto2015generalized} and this suggests the dual $\Z_q^{(d-2)}$ symmetry of $\T/\Z_q$ is broken by the mixed anomaly \eqref{U(1) grp ext anom} after gauging also $\wt\U(1)$.\footnote{This can be seen from \eqref{wilson line A Atilde S}: if we gauge $\wt A\to\wt a$, the Wilson lines of $s$ are always dressed with the ones of $\wt a$ and absorbed in those of $a$. This shows that the $\Z_q^{(d-2)}$ symmetry is broken (the Wilson lines are not topological anymore). Notice the difference with the discrete case. There $S$ should be dressed like \eqref{wilson line A Atilde S}, but the dressing is still given by another discrete gauge field such that the final result is again topological. So, in spite of the anomaly, after gauging we can still combine the objects to create a topological Wilson line, generator of $\wh G^{(d-2)}$. It follows that $\wh A^{(d-2)}$ is instead preserved somehow to fit in $\wh G^{(d-2)}$.} However, this is not the whole story and we will be missing some information by just setting $C =0$.

Consider the case with $C=0$. Gauging $\wt A\to \wt a$ gives the partition function
\begin{equation}
\label{gauging U(1)tilde C=0}
    \sum_{s|\d s =c_1(\wt a)|_q}\int \mathcal{D} \wt a \,Z[\wt a,s] e^{ 2\pi i \int_Y \frac{\d \wt a}{2\pi} \wedge \frac{\d \wt B_m}{2\pi} }, 
\end{equation}
with the standard magnetic coupling to $\U(1)_m^{(d-3)}$ \cite{gaiotto2015generalized}, with background $B_m$,\footnote{In obvious notation, $\U(1)_m^{(d-3)}$ is the dual symmetry of $\U(1)$ with background $B_m$, while $\wt \U(1)_m^{(d-3)}$ is the magnetic symmetry of $\wt \U(1)$ with background $\wt B_m$.} 
\begin{equation}
\label{Bm coupling}
    i\int_X \frac{\d a}{2\pi}B_m=2\pi i\int_Y \frac{\d a}{2\pi} \wedge \frac{\d B_m}{2\pi}, \qquad \p Y = X.
\end{equation}
In \eqref{gauging U(1)tilde C=0} there is also implicit a Maxwell kinetic term for $\wt a$. The partition function \eqref{gauging U(1)tilde C=0} is well-defined and gauge invariant under \eqref{extra gauge transf for U(1) ext}. In particular, the magnetic coupling can be rewritten, using the identification $qa = \wt a$, as 
\begin{equation}
    2\pi i\int_Y \frac{\d a}{2\pi} \wedge q\frac{\d \wt B_m}{2\pi} = 2\pi i\int_Y \frac{\d a}{2\pi} \wedge \frac{\d B_m}{2\pi}
\end{equation}
with the identification 
\begin{equation}
\label{Bm=Bmtilde C=0}
    B_m = q\wt B_m.
\end{equation}
This is fine, but it is clearly missing some information, since this would imply that the fluxes of $B_m$ are multiple of $q$, which is generically not true. One way out from this is to say that, since $\wt a$ has fluxes multiple of $q$ that couple to $\wt B_m$ as in \eqref{Bm coupling}, then the coupling is gauge invariant even if we allow $\wt B_m$ to have fluxes
\begin{equation}
\label{tildeBm fractional fluxes}
    \oint \frac{\d \wt B_m}{2\pi} \in \frac{1}{q}\Z,
\end{equation}
so that $B_m$ has the standard quantization. This is morally true, but a better way to encode the information \eqref{tildeBm fractional fluxes} is to consider also a $\Z_q$ gauge field that reproduces the fractional $1/q$ fluxes. This will be the role of $C$.

To see how this works properly, it is convenient to consider flat gauge fields, which simplifies the analysis. Indeed, we are looking for mod $q$ effects in the fluxes of $B_m$ that are not seen in a standard differential form approach. Moreover, as already noticed, the non-trivial content of the extension \eqref{U(1) grp ext} is in its topological part, so that $A$ and $\wt A$ can be taken to be flat. The kinetic Maxwell term does not plat a role in the analysis, since, using the local identification $qA = \wt A$, we have that
\begin{equation}
    \frac{1}{2q^2} \wt f \wedge *\wt f=  \frac{1}{2} f \wedge *f,
\end{equation}
where $q$ is the minimal charge of $\wt a$, having normalized the fundamental charge of $a$ to one.

\subsection{Flat gauging}

To adopt a uniform notation for both discrete and continuous fields, it is convenient here to view $\U(1)\simeq\R/\Z$. The sequence \eqref{U(1) grp ext} is equivalent to
\begin{equation}
\label{R/Z grp ext}
    \Z_q \xrightarrow{1/q}  \R/\Z \xrightarrow{\cdot q} \wt \R/\Z, \qquad [\a] \in H^2(B\wt\U(1);\Z_q) \simeq \Z_q.
\end{equation} 
Here $A \in H^1(X;\R/\Z)$ and the condition \eqref{U(1) ext gauge fields} is\footnote{The first Chern class for flat $A$ is $\b'(A)=c_1(A)$, for $\b': H^1(X;\R/\Z)\to H^2(X;\Z)$ the Bockstein for the long exact sequence induced by $\Z \to \R \to \R/\Z$. In \eqref{R/Z ext gauge fields}, $\b(\wt A)=\b'(\wt A) \mod q$.}
\begin{equation}
\label{R/Z ext gauge fields}
    \d S= \wt A^*\a = \b(\wt A) = c_1(\wt A)|_q,
\end{equation}
where $\b$ is the Bockstein of the long exact sequence induced by \eqref{R/Z grp ext}, i.e.
\begin{equation}
    ... \to H^1(X;\wt\R/\Z)\xrightarrow{\b} H^2(X;\Z_q) \to H^2(X;\R/\Z) \to H^2(X;\wt\R/\Z) \to...
\end{equation}
The gauge field $A$ can be decomposed with the pair $(S,\wt A)$ as\footnote{With a little abuse of notation, $\wt A' = 1/q \wt A$. The decomposition \eqref{flat A with Atilde S} is the same as \eqref{wilson line A Atilde S}.}
\begin{equation}
\label{flat A with Atilde S}
    A = \wt A' - \frac{1}{q} S,
\end{equation}
for a lift of $\wt A \in H^1(X;\wt \R/\Z)$ to $\wt A' \in C^1(X;\R/\Z)$, such that $\d \wt A' = 1/q \b(\wt A)$. The condition \eqref{U(1) ext gauge fields} ensures that $A$ is flat. The gauge transformations \eqref{extra gauge transf for U(1) ext} are
\begin{equation}
\label{gauge transf for R/Z ext}
    \wt A \to \wt A +\d \wt \l,\qquad \b(\wt A) \to \b(\wt A) +\d \w, \qquad S\to S+\d \g+ \w, \qquad \w \in C^1(X;\Z_q).
\end{equation}
The gauge transformation for $\b(\wt A)$ comes from the ambiguity of the lift $\wt A \to \wt A'$, 
\begin{equation}
    \wt A' \to \wt A' + \d \wt \l '+ \frac{1}{q}\w.
\end{equation}

Gauging $\U(1)$ produces a theory $\T/\U(1)$ with a dual $\Z^{(d-2)}\simeq \hom(\R/\Z,\R/\Z)$ magnetic symmetry. This can be identified with the topological part of the more familiar magnetic symmetry and we can thus identify its background as the generalized Chern class of the would-be $B_m$, i.e. $c_1(B_m) \in H^{d-1}(X;\Z)$. We now do the same in two steps. Gauging $\Z_q$ gives a theory with dual symmetry $\Z_q^{(d-2)}$, with background $C$, and mixed anomaly \eqref{U(1) grp ext anom}, 
\begin{equation}
\label{R/Z grp ext anom}
        \frac{2\pi i}{q} \int_Y \b(\wt A) \cup C.
\end{equation}
Next we gauge $\wt \U(1)$, by first adding a counterterm $2\pi i \,\wt A'\cup \wt C$,\footnote{Notice that it is well-defined, since $\wt A'$ is $\R/\Z$ valued and $\wt C$ is $\widehat{\R/\Z} \simeq \Z$ valued. This is the same counterterm found in \eqref{f for Znm}.} and then summing over $\wt A$ with a coupling to the dual background $c_1(\wt B_m)$ for $\wt \Z^{(d-2)}$, 
\begin{equation}
\label{Z gauge Zq R/Ztilde}
    \sum_{\wt a, s|\d s= \b(\wt a)} Z[s,\wt a]e^{\int_X \frac{2\pi i}{q}s\cup C - 2\pi i\, \wt a'\cup \wt C+ 2\pi i \,\wt a \cup c_1(\wt B_m)}.
\end{equation}
One can check that this partition function is gauge invariant under \eqref{gauge transf for R/Z ext} provided that
\begin{equation}
\label{gauge field dual sym of R/Z ext}
    \d c_1(\wt B_m) = \wh \b(C), \qquad \wh \b: H^{d-1}(X;\Z_q) \to H^d(X;\Z),
\end{equation}
where $\wh \b$ is the Bockstein for the sequence $\Z \to \Z \to \Z_q$. This shows that the anomaly \eqref{R/Z grp ext anom} does not prevent to gauge $\wt \U(1)$, but imposes the constraint \eqref{gauge field dual sym of R/Z ext} on the dual symmetries $\Z_q^{(d-2)}$ and $\wt \Z^{(d-2)}$, which are not independent symmetries but fit in the extension
\begin{equation}
\label{dual sym of R/Z ext}
    \wt \Z^{(d-2)} \to \Z^{(d-2)} \to \Z_q^{(d-2)},
\end{equation}
according to \eqref{gauge field dual sym of R/Z ext}. The full background for $\Z^{(d-2)}$ is thus given by
\begin{equation}
    \label{Bm = Bmtilde + C}
    c_1(B_m) = qc_1(\wt B_m) - \wt C, \qquad \d c_1(B_m)=0.
\end{equation}
In particular 
\begin{equation}
    c_1(B_m) = -C \mod q.
\end{equation}
This is the improved version of \eqref{Bm=Bmtilde C=0}. Locally $B_m=q\wt B_m$ and the fractional fluxes \eqref{tildeBm fractional fluxes} are taken into account by $C$. 

To be more explicit, the partition function \eqref{Z gauge Zq R/Ztilde} of $\T/\Z_q/\wt \U(1)$ can be rewritten, using \eqref{flat A with Atilde S} and so $q a = \wt a - s$, as
\begin{equation}
\label{Z gauge R/Z}
\begin{split}
    &\sum_{\wt a, s|\d s= \b(\wt a)} Z[s,\wt a]e^{2\pi i \int_X -(\wt a'-\frac{1}{q}s)\cup \wt C +\wt a \cup c_1(\wt B_m)} =          \\
    &\sum_{\wt a, s|\d s= \b(\wt a)} Z[s,\wt a]e^{2\pi i \int_X - a \cup \wt C + a \cup qc_1(\wt B_m)}  =  \sum_{a} Z[a]e^{2\pi i \int_X a \cup (qc_1(\wt B_m)-\wt C)}.
\end{split} 
\end{equation}
This gives again the identification \eqref{Bm = Bmtilde + C} and shows that $\T/\U(1)\simeq \T/\Z_q/\wt \U(1)$.

We conclude this section with some comments. Since we restricted to flat gauge fields, the discussion is analogous to the finite groups case. In particular, the dual symmetry \eqref{dual sym of R/Z ext} is expected on general grounds, being the Pontryagin dual sequence of \eqref{R/Z grp ext} (or \eqref{U(1) grp ext}). The Bockstein $\wh \b$ in \eqref{gauge field dual sym of R/Z ext} is indeed the cohomological operation that classifies \eqref{dual sym of R/Z ext}, which is the dual operation of $\b$ in \eqref{R/Z ext gauge fields} on a $d+1$ dimensional $Y$ according to \eqref{hat alpha from alpha}:\footnote{However, notice that the sequence \eqref{R/Z grp ext} is classified by $H^2(K(\Z,2);\Z_q)\simeq \hom(\Z,\Z_q)$ which is not $\b$. So $\wh \b\in H^{d}(K(\Z_q,d-1),\Z)$ classifies \eqref{dual sym of R/Z ext} and it is the same as in the finite group case, while $\b$ is not.}
\begin{equation}
    \b :  H^1(-;\R/\Z) \to H^2(-;\Z_q), \qquad
    \wh \b: H^{d-1}(-;\wh \Z_q)\to H^{d}(-;\widehat{\R/\Z}), 
\end{equation}
where $\wh \Z \simeq \R/\Z$. Indeed, the anomaly \eqref{R/Z grp ext anom} can be easily rewritten as in \eqref{f for Znm},\footnote{A more symmetric form can be achieved by using $C$ valued in $\wh \Z_q$, so that the first term has not the $1/q$ factor.}
\begin{equation}
\label{R/Z grp ext anom dual}
        \frac{2\pi i}{q} \int_Y \b(\wt A) \cup C = 2\pi i \int_Y \wt A\cup \wh\b(C)+2\pi i\int_X \wt A'\cup \wt C, \qquad \p Y= X.
\end{equation}
On closed $Y$, the two expressions with $\b$ and $\wh \b$ define the same SPT phase (anomaly), whose difference is just a local counterterm on $X$ (which is the one we added in \eqref{Z gauge Zq R/Ztilde}). 

Finally, one could check that the dual symmetry \eqref{dual sym of R/Z ext} is not anomalous. The partition function of $\T/\Z_q/\wt \U(1)$ is invariant under the background gauge transformations of \eqref{gauge field dual sym of R/Z ext}, namely
\begin{equation}
    C\to C+\d \wh \g, \qquad \wt C\to \wt C+\d\widetilde{\wh \g}+q \wh \w, \qquad c_1(\wt B_m)\to c_1(\wt B_m)+ \wh\w, 
\end{equation}
provided that \eqref{R/Z ext gauge fields} holds.

\subsection{Dynamical gauging} 

We argued that \eqref{U(1) grp ext} affects only the topological part of the $\U(1)$ symmetry and therefore the flat gauging analysis in the previous section should suffice. However, one could worry that when the photon dynamics is considered (dynamical gauging), the result \eqref{Bm = Bmtilde + C} can be modified. We show here that this is not the case: namely, the identification \eqref{Bm = Bmtilde + C} still holds, but now of course they are not backgrounds for an independent $\Z^{(d-2)}$ symmetry but the topological data of the magnetic $\U(1)_m^{(d-3)}$ symmetry. 

Since we should carefully taking into account the topological part of the electromagnetic field, it is convenient to use a differential cohomology approach \cite{CheegerSimons1985,HopkinsSinger2002DiffCo} (for reviews \cite{tachikawa2020pformanom,gwmooreSaxenaTASITQFT,DEBRAY2025DiffCoReview}).\footnote{The original mathematical papers \cite{CheegerSimons1985,HopkinsSinger2002DiffCo} are clear and quite readable also for physicists.} This is needed for two reasons. First, from \eqref{Bm = Bmtilde + C} we see that the discussion is sensible to torsion, not detected usually in the differential form approach. Secondly, the anomaly itself \eqref{U(1) grp ext anom} requires to consider the gauge transformation $c_1(A)\to c_1(A)+\d n$, which is again not taken into account when we identify $c_1(A)$ with $F/2\pi$.\footnote{Written naively with continuous gauge fields, so $c_1(\wt A)= \wt F/2\pi$ and $C \to q/2\pi C$, the anomaly is actually trivial for $\wt \U(1)$, 
\[ \int_X \frac{i}{2\pi} F \wedge C.\] So the differential form language is missing some piece of information.}

\paragraph{Differential cohomology: basics.} The idea is quite simple. The gauge invariant data of an $\U(1)$ $p$-form gauge field are its field strength $F$, a closed $p+1$ form, and its holonomies $\chi: Z_p(X;\Z)\to \U(1)$, sending $\S \mapsto \chi(\S)$ for a closed $p$-dimensional submanifold $\S$. They are related by the condition that, if $\p V = \S$, then 
\begin{equation}
\label{differential character}
    \chi(\S=\p V)= e^{2\pi i\int_V F}.
\end{equation}
Such an object is called a differential character \cite{dijkgraafwitten,CheegerSimons1985} and the set of all such objects is called $\check{H}^{p+1}(X;\Z)\coloneqq \check{H}^{p+1}(X)$,\footnote{Compared to the original paper \cite{CheegerSimons1985} there is a shift by one in the degree. This is the notation of \cite{HopkinsSinger2002DiffCo}.} the $(p+1)$-th differential cohomology group of $X$. These data can be reformulated by introducing $A \in C^p(X;\R)$ such that $\chi(\S)=\exp(2\pi i\int_\S A)$. In the extension to $A$ there is some gauge freedom given by $A \to A+\d \l +n$: $\d\l$, $\l\in C^{p-1}(X;\R)$, because originally $\chi$ is defined only for closed spaces $\S$; $n\in C^p(X;\Z)$ because it gives $\chi = 1$. Moreover, because of \eqref{differential character}, it must exist $N \in Z^{p+1}(X;\Z)$ such that $F-N = \d A$. $N$ is the integer flux of $A$, i.e. its generalized Chern class, so $N =c_1(A)=c_1$. Notice that under $A\to A+\d n$ it transforms as $c_1 \to c_1-\d n$, so only $[c_1]\in H^{p+1}(X;\Z)$ matters. All together, the triple $\check{A}=(c_1,A,F)$ is a $p+1$ differential cocycle whose gauge invariant class corresponds to an element of $\check{H}^{p+1}(X)$.

The benefit of this approach is both to take into account and separate clearly the topological data, namely $c_1$, and the geometrical ones, like $F$.\footnote{Under the inclusion map $\N \xhookrightarrow{} \R$, $[c_1]_\R=[F]$. This is the standard identification modulo $2\pi$ because of the normalization of $A$ used here.} Notice that when $\d n=0$, $c_1$ is invariant under gauge transformations and $A\to A+n$ is basically the ordinary notion of large gauge transformation. For a flat gauge field, $F=0$, it is easy to show that the differential cocycle $(c_1,A,0)$ corresponds to an element of $H^p(X;\R/\Z)$, since $\d A = c_1$, so $\d A =0$ in $\R/\Z$. The converse is also true: starting from $A'\in H^p(X;\R/\Z)$, one can construct a differential cocycle by lifting $A'$ to $\R$ and $c_1= \b(A')$. Therefore, there is an isomorphism $\check{H}_{\rm flat}^{p+1}(X)\simeq H^p(X;\R/\Z)$. The description of the previous section is thus automatically encoded in the differential cohomology point of view.

Take two differential cocycles $\check{A}\in \check{H}^{p+1}(X)$ and $\check{A}'\in \check{H}^{q+1}(X)$. Their product $\check{A}\star \check{A}' \in \check{H}^{p+q+2}(X)$ is defined by requiring that for the flux and the curvature it reproduces the standard notions, namely $c_1\cup c_1'$ and $F\wedge F'$. This yields for the gauge fields 
\begin{equation}
    \label{star product}
    A \star A' = A\cup c_1'+ (-1)^{p+1} F\cup A'+ Q(F,F'),
\end{equation}
where $Q(F,F')$ is a correction term that measures the difference between $F \wedge F$ and $F \cup F$ after embedding $F$ as an $\R$-valued cochain \cite{tachikawa2020pformanom}.\footnote{Namely $a\wedge b- a \cup b = Q(\d a, b)+(-1)^{|a|}Q(a,\d b)+\d Q(a,b)$. $Q$ is a cochain homotopy that gives the difference between two different definitions of cup product on cochains (here $\cup$ and the one induced by $\wedge$) that descent to the same cup product in cohomology.} This product is well-defined on cohomology classes, in the sense that gauge transformations of $\check{A}$ or $\check{A}'$ leave $\check{A}\star \check{A}'$ invariant up to gauge transformations. One can check that $\d (A\star A')=F\wedge F'-c_1\cup c_1'$. Notice that, if $\check{A}\star \check{A}' \in \check{H}^{d+1}(X)$, then this is necessarily flat, with $\d(A\star A')=0$, i.e. $A\star A'\in H^d(X;\R)$, which can be naturally integrated over $X$. It follows that the natural coupling between an electromagnetic gauge field $\check{a}=(c_1(a),a, f) \in \check{H}^{2}(X)$ and its magnetic background $\check{B}_m=(c_1(B_m),B_m,F_m)\in \check{H}^{d-1}(X)$ is given by \cite{tachikawa2020pformanom}
\begin{equation}
    \label{Bm coupling diff coho}
    S \supset 2\pi i\int_X a \star B_m = 2\pi i \int_X a\cup c_1(B_m)+ f\cup B_m+ Q(f,F_m).
\end{equation}
This is invariant under gauge transformation of $a$ and magnetic background transformations of $B_m$. Notice how \eqref{Bm coupling diff coho} reduces correctly to the standard differential form coupling \eqref{Bm coupling} in the topologically trivial case, $c_1(a)=0$,\footnote{Indeed, for $c_1(a)=0$, $a\star B_m = a\wedge F_m+ \d(...)$, i.e. \eqref{Bm coupling}. This can be shown using $Q(f=\d a,F_m)=a \wedge F_m - a \cup F_m-\d Q(a,F_m)$.} and to the flat case \eqref{Z gauge Zq R/Ztilde} when $f=0$, interpolating between the two.

We have directly introduced the differential cohomology groups: physically, they arise naturally from the gauge invariant data of a gauge field. However, it is also possible to introduce the notion of differential cochains with a suitable differential map. A differential cochain is a triple $(c,a, f)\in C^{p+1}(X;\Z)\times C^p(X;\R)\times \W^{p+1}(X)$ and the Hopkins-Singer differential $\d_{\rm HS}$ acts on them as \cite{HopkinsSinger2002DiffCo}
\begin{equation}
    \d_{\rm HS}(c,a,f)\coloneqq (\d c, c-f -\d a, \d f).
\end{equation}
One can check that $\d_{\rm HS}^2=0$. A differential cocycle is a differential cochain closed under $\d_{\rm HS}$.

\paragraph{$\U(1)$ extension.} We can now go back to the question of gauging \eqref{U(1) grp ext}. The main point that we learn from the excursus on differential cohomology is the correct form of the coupling to the magnetic background \eqref{Bm coupling diff coho} in full generality. Taking this into account, the discussion is pretty much equivalent to the flat case. Locally $qf =\wt f$ and $a=1/q \,\wt a-1/q\,\wt s$ \eqref{flat A with Atilde S}.\footnote{So $\d q a = qf -qc_1=\d \wt a -\d \wt s= \wt f -(\wt c_1+\d \wt s)$, which gives $qf =\wt f$, $qc_1 = \wt c_1+\d \wt s$.} The partition function of the gauged theory $\T/\Z_q/\wt \U(1)$ is
\begin{equation}
    \sum_{\wt a, s|\d s= c_1(\wt a)|_q} Z[s,\wt a]e^{2\pi i\int_X \frac{1}{q}s\cup C - \frac{1}{q}\wt a\cup \wt C+\wt a \cup c_1(\wt B_m)+\wt f \cup \wt B_m+Q(\wt f,\wt F_m) +\int_X \frac{1}{2q^2}\wt f\wedge *\wt f},
\end{equation}
with the counterterm as in \eqref{Z gauge Zq R/Ztilde}. Gauge invariance under $\wt a \to \wt a +\d \wt\l+\wt n$, $\wt s \to \wt s +\wt n$ requires that $\d c_1(\wt B_m) =\wh \b(C)$, i.e. \eqref{gauge field dual sym of R/Z ext}. Rewritten in terms of $a$, it becomes
\begin{equation}
    \sum_{a} Z[a]e^{2\pi i\int_X -a \cup \wt C+a \cup q c_1(\wt B_m)+ f \cup q\wt B_m+qQ( f,\wt F_m) +\int_X \frac{1}{2}f\wedge * f}.
\end{equation}
This is again equivalent to gauge $a$ with the magnetic coupling $a\star B_m$ with the identification $qc_1(\wt B_m)-\wt C=c_1(B_m)$ \eqref{Bm = Bmtilde + C} and locally $q\wt B_m =B_m$ ($q\wt F_m=F_m$).

These are all the results already obtained with the flat gauging, but the interpretation is a little different: there is no independent symmetry as \eqref{dual sym of R/Z ext} in the final theory, but this should be viewed as the topological part involving the generalized Chern class of the magnetic $\U(1)_m^{(d-3)}$ symmetry. We can formalize this argument even further. There is an isomorphism $H^p(K(\Z_q,n);\R/\Z)\simeq H^{p+1}(K(\Z_q,n);\Z)$.\footnote{For every $X$ with $H^p(X;\R)=0$, $H^p(X;\R/\Z)\simeq H^{p+1}(X;\Z)$, because of the exact sequence $\Z \to \R \to \R/\Z$. $K(G,n)$ for a finite Abelian $G$ has only torsion homologies \cite{FomenkoFuchs2016}, therefore $H^p(K(G,n);\R)=0$.} The dual cohomology class $\wh \b\in H^d(K(\Z_q,d-1);\Z)$ is thus specified by $\wh \a \in H^{d-1}(K(\Z_q,d-1);\R/\Z)\simeq \hom(\Z_q,\R/\Z)$, such that $\wh \b = \b' \cdot \wh \a$, with $\b'$ the Bockstein for $\Z \to \R \to \R/\Z$ ($\wh \a$ is basically the multiplication by $1/q$). Notice that $\wh \a$ is exactly the dual cohomological operation \eqref{hat alpha from alpha} applied to \eqref{U(1) grp ext}. Using $\wh \a$, it is possible to construct $\check{C}=\wh \a(C)\in H^{d-1}(X;\R/\Z)\simeq \check{H}^{d}_{\rm flat}(X)$, such that $\check{C}=(\b( C)=\b'(\wh\a(C)),\wh \a(\wt C),0)$, with $\wt C$ a $\Z$-lift of $C$ (so $\wh \a (\wt C)$ is an $\R$-valued cochain). The relation \eqref{Bm = Bmtilde + C} can be lifted to the differential cochains as
\begin{equation}
    \label{Btilde C diff coho}
    \d_{\rm HS}\check{\wt B}_m = \check{C} =\wh\a(C),
\end{equation}
which contains also the condition\footnote{Notice that this is consistent with the identification used above: $c_1(B_m)=qc_1(\wt B_m)-\wt C$, $B_m=q\wt B_m$, $F_m=q\wt F_m$ (since $\wh \a= 1/q\cdot$).}
\begin{equation}
    \d \wt B_m= \wt F_m - c_1(\wt B_m)-\wh \a (\wt C).
\end{equation}
This says that under the gauge transformation $C\to C +\d \g$, also $\wt B_m$ transforms as $\wt B_m\to \wt B_m -\wh\a(\wt \g)$ (and $c_1(\wt B_m)\to c_1(\wt B_m)-\wt n$ under $\wt C\to \wt C +\wt n$, but this is also in \eqref{Bm = Bmtilde + C}). All together, these results implies that dual symmetry after gauging \eqref{U(1) grp ext} is the extension
\begin{equation}
    \label{dual U1 extension diff coho}
    \wt \U(1)_m^{(d-3)}\to \U(1)_m^{(d-3)}\to \Z_q^{(d-2)}, \qquad [\wh \a] \in H^{d-1}(K(\Z_q,d-1);\R/\Z),
\end{equation}
which gives the condition \eqref{Btilde C diff coho}. 

The final result is that, by suitable formulating the $\U(1)$ gauge fields as differential characters, which is the natural way to encode all their topological and geometric data, the $\U(1)$ case is exactly the same to the discussion with discrete groups \eqref{grp ext higher-form} (besides a shift by one in the degrees due to $B\U(1)\simeq K(\Z,2)$). \\

\noindent\textit{Note.} We have not mentioned the $\th$-term for the $\U(1)$ gauge field, but this can be included trivially similarly to the Maxwell kinetic term without spoiling the discussion. For example in four dimensions 
\begin{equation}
   \frac{\wt\th}{4\pi^2}\int \wt f \wedge \wt f = \frac{\wt \th q^2}{4\pi^2}\int f \wedge f.  
\end{equation}
The theta angles are related by $\th = \wt\th q^2$. In fact, because fluxes of $\wt a$ are valued in $q\Z$ (since $qc_1(a) = c_1(\wt a)$), the periodicity of $\wt \th$ is $\wt \th \sim \wt \th + 2\pi/q^2$. This is again due to the fact that the minimal electric charge for $\wt \U(1)$ is $q$ times the one for $\U(1)$.

\subsection{Example: mixed magnetic anomalies}

Consider a theory $\T$ with the following symmetry
\begin{equation}
\label{mixed U1 G group}
    \frac{\U(1)\times G}{\Z_q},
\end{equation}
for some group $G$ (which we can also take to be a spacetime symmetry group), with $\Z_q\subset Z(G)$. This could be realized by an extension of the form
\begin{equation}
\label{ext for magnetic anom}
    \Z_q\to \frac{\U(1)\times G}{\Z_q}\to \wt \U(1)\times \wt G,
\end{equation}
given by $(u,g)\mapsto (u^q,\pi(g))$, where $\pi: G\to \wt G\simeq G/\Z_q$. The extension \eqref{ext for magnetic anom} is specified by a class $[\a]\in H^2(B(\wt U(1)\times \wt G); \Z_q)$. Using that $B(\wt \U(1)\times \wt G)\simeq B\wt \U(1)\times B\wt G$,\footnote{Take two groups $G$ and $H$. Start with $EG\times EH$, where $EG$ ($EH$) is a contractible space with $G$ ($H$) action, and define the $G \times H$ action trivially on each single factor. Then $EG\times EH/ (G\times H) \simeq EG/G \times EH/H\simeq BG \times BH$.} the K\"unneth theorem says that \cite{HatcherAT,FomenkoFuchs2016}
\begin{equation}
\label{kunneth}
    H^2(B(\wt \U(1)\times \wt G);\Z_q) \simeq H^2(B\wt\U(1);\Z_q)\oplus H^2(B\wt G;\Z_q),
\end{equation}
since $H^1(B\U(1);\Z_q)=0$ and it generically has no torsion. Thus $\a$ splits as 
\begin{equation}
\label{alpha splits kunneth}
\a = \a(\tilde \U(1))+\a (\wt G),
\end{equation}
where $\a(\U(1))=c_1(\U(1))$ mod $q$ and $\a(\wt G)\in H^2(B\wt G;\Z_q)$ classifies the extensions $\Z_q\to G \to \wt G$.

Now, we can gauge the $\U(1)$ factor in \eqref{mixed U1 G group} to obtain a new theory $\T/\U(1)$ with a dynamical $\U(1)$ connection $a$ and the consequent magnetic dual symmetry $\U(1)_m^{(d-3)}$. Notice that, after gauging $\U(1)$, the zero-form global symmetry is $\wt G$, since $\Z_q\subset G$ is part of the gauge group. Because of the quotient in \eqref{mixed U1 G group}, the fluxes of $a$ must satisfy 
\begin{equation}
    [c_1(\wt a)] =  -\G^*[\a(\wt G)] \mod q,   
\end{equation}
with $\G:X\to B\wt G$ the background field for $\wt G$. This is the vanishing of the pullback of \eqref{alpha splits kunneth} and, in terms of $qc_1(a)=c_1(\wt a)$, it tells us that $a$ could have fractional fluxes $1/q \G^*\a(\wt G)$. It follows that the usual magnetic coupling \eqref{Bm coupling} implies a mixed anomaly
\begin{equation}
\label{mixed magnetic anomaly}
    -\frac{2\pi i}{q}\int_Y \G^*\a(\wt G)\cup \frac{\d B_m}{2\pi},
\end{equation}
as follows by checking the independence from the extension of the correct magnetic coupling \eqref{Bm coupling}.\footnote{Without the extension, that it is not always possible, the anomaly \eqref{mixed magnetic anomaly} can be checked using the refined coupling $2\pi i\int a\star B_m$ \eqref{Bm coupling diff coho}. This is not invariant under background gauge transformations of $B_m$ because of the fractional fluxes $c_1(a)=1/q\, \G^*\a(\wt G)$. This gives exactly the anomaly \eqref{mixed magnetic anomaly}, with $\d B_m/2\pi \to c_1(B_m)$.} This is a general consequence for the magnetic symmetry of a $\U(1)$ gauge group that fits in \eqref{mixed U1 G group}.

We will now see that the result of the previous section is consistent with \eqref{mixed magnetic anomaly} and, in fact, the mixed anomaly \eqref{mixed magnetic anomaly} is a consequence of \eqref{group extension anomaly app}. As shown above, gauging $\U(1)$ in \eqref{mixed U1 G group} is equivalent to first gauge $\Z_q$ and then $\wt \U(1)$. The background field for $\Z_q$ is the object that trivializes the pullback to $X$ of \eqref{alpha splits kunneth} and gauging $\Z_q$ yields a theory $\T/\Z_q$ with symmetry $\wt \U(1)\times \wt G\times \wh \Z_q^{(d-2)}$ and mixed anomaly 
\begin{equation}
\label{mixed anomaly U1 G Z}
     \frac{2\pi i }{q} \int_Y (c_1(\wt A) + \G^* \a(\wt G)) \cup C, 
\end{equation}
with $C$ the background for $\wh \Z_q^{(d-2)}$. The first term is the same anomaly as \eqref{U(1) grp ext anom} and, after gauging $\wt \U(1)$, it is the piece of information that gives the result \eqref{Bm = Bmtilde + C} so that the final theory $\T/\Z_q/\wt \U(1)$ has $c_1(B_m) = -C \mod q$. The second term remains as an anomaly and produces \eqref{mixed magnetic anomaly} after setting $c_1(B_m) = -C \mod q$.

\subsection{Generalizations}
 
We have shown that, after gauging the extension \eqref{U(1) grp ext}, the dual symmetry is the extension \eqref{dual U1 extension diff coho}, defined by \eqref{Btilde C diff coho} (which are the refined versions of \eqref{dual sym of R/Z ext} and \eqref{gauge field dual sym of R/Z ext}). We can also consider higher-form analogs of \eqref{U(1) grp ext}, similar to what done for finite groups. 

For a $\U(1)^{(p)}$ symmetry the classifying space is $K(\Z,p+2)$, whose universal cohomology class in $H^{p+2}(K(\Z,p+2);\Z)\simeq \Z$ gives the generalized Chern class $c_1\in H^{p+2}(X;\Z)$. The extensions of a $\U(1)^{(p)}$ symmetry by a $\Z_q^{(l)}$ symmetry are classified by $H^{l+2}(K(\Z,p+2);\Z_q)$. The case $p=l$, the higher-form extension $\Z_q^{(p)}\to \U(1)^{(p)}\to \U(1)^{(p)}$, is exactly the same as \eqref{U(1) grp ext} (i.e. the $p=l=0$ case). After gauging $\U(1)^{(p)}$, the background $B_m$ for the dual magnetic symmetry $\U(1)_m^{(d-p-3)}$ satisfies \eqref{Bm = Bmtilde + C}. This implies the dual extension 
\begin{equation}
    \U(1)_m^{(d-p-3)}\to \U(1)_m^{(d-p-3)}\to \Z_q^{(d-p-2)},
\end{equation}
as in \eqref{dual U1 extension diff coho}, resulting in the condition \eqref{Btilde C diff coho}, i.e. 
\begin{equation}
    \d_{\rm HS}\check{B}_m=\check{C}= \wh \a(C),
\end{equation}
using again $\wh \a(C)\in H^{d-p-1} (X;\R/\Z)\simeq\check{H}_{\rm flat}^{d-p}(X)$, with $C$ the background for $\Z_q^{(d-p-2)}$. Without recurring to differential characters, this could be understood from a flat gauging perspective, mirroring \eqref{Bm = Bmtilde + C}. Using the isomorphism $H^{d-p-1}(K(\Z_q,d-p-1);\R/\Z)\simeq H^{d-p}(K(\Z_q,d-p-1);\Z)$, given by $\b'\wh \a=\wh \b$, we obtain $\d c_1(B_m)=\wh \b(C)$. 

It is clear from our discussion that to describe on the same footing and more conveniently extensions (or higher-groups) of both $\U(1)$ and discrete symmetries (possibly with multiple $\U(1)$ factors of different degrees), it would be nice to have a systematic theory of natural differential cohomological operations (see for example \cite{GradySati2018}).

\section{Extensions and fractionalization}
\label{sec symfrac}

The result of gauging $A$ in \eqref{general grp ext} is related to symmetry fractionalization \cite{barkeshli2014symfrac,KomargodskiHsinSymFrac2022,DumitrescuCordovaBrennanSymFrac2022,HsinShao2020,brennanJacobsonRoumpedakis2025SymFrac} and it can be described with  the language of cohomological operations introduced before. This is implicit in  \cite{tachikawaGaugeFiniteGroups} and better emphasized in \cite{brennanJacobsonRoumpedakis2025SymFrac}. See also Appendix \ref{app sym frac} for a review of symmetry fractionalization.

Consider the central extension \eqref{general grp ext}, where now we could allow $K$ to be an arbitrary group (while $A$ is always finite and Abelian, so that $\wh A \simeq A$). After gauging $A$, the Wilson lines $W(\g)$ for the $a$ gauge field are not quite topological in presence of a non-zero background $K$, because of the general condition $\d a = K^*\a$ \eqref{gauge field grp ext}.\footnote{In this section we go back to denote the gauge fields with the same letter of their groups.} This means that they have a surface dependence given by
\begin{equation}
\label{surface dependence wilson line ext}
    e^{2\pi i\int_\S K^*\a}, \qquad \p \S= \g.
\end{equation}
This is of course the anomaly \eqref{group extension anomaly app} in terms of  symmetry defects, since the Wilson lines of $a$ are the generators of the dual $\wh A^{(d-2)}$ symmetry. Indeed, by Poincaré duality (PD), inserting $W(\g)$ in the correlators is equivalent to turning on a background field $\wh A = \rm PD(\g)$; therefore
\begin{equation}
\label{anomaly sym frac is grp ext anom}
    Z[\wh A+\d \l] = \braket{W(\g+\p\S)} = \braket{W(\g)}e^{2\pi i\int_\S K^*\a} =  Z[\wh A]e^{2\pi i\int_X K^*\a \cup \l}, 
\end{equation}
where $\l$ is the Poincaré dual of $\S$. This is the anomaly of \eqref{group extension anomaly app}.

The surface dependence \eqref{surface dependence wilson line ext} is a kind of anomaly inflow for the quantum mechanical theory living on the line $\g =\p\S$ (if we think of the Wilson lines as arising from tracing out high-energy degrees of freedom). Anomalies in quantum mechanics (i.e. $0+1d$ QFT) for a symmetry $K$ are given by its projective representations classified by $H^2(BK;\U(1))$. The condition \eqref{surface dependence wilson line ext} is not exactly the same, since $\a \in H^2(BK; A)$, but it often happens that in fact $H^2(BK;\U(1))\simeq H^2(BK;A)$ for some cyclic group $A$ and $K$ a compact connected Lie group.\footnote{Projective representations of some connected $G$ come from linear representations of the universal covering $\wt G$, $G\simeq \wt G/\pi_1(G)$, when $\chi: \pi_1(G) \to \U(1)$ is non-trivial ($\pi_i(BG)=\pi_{i-1}(G)$ from the long exact sequence in homotopy and therefore for $G$ connected $\pi_1(BG)\simeq H_1(BG)=0$ and $\pi_2(BG)\simeq H_2(BG)\simeq\pi_1(G)$; therefore  $H^2(BG;\U(1))\simeq \hom(H_2(BG),\U(1))\simeq \hom(\pi_1(G),\U(1))$). For usual compact connected Lie groups, except $\U(1)$, $\pi_1(G)$ is just torsion, i.e. some cyclic factor, and therefore projective representations come from considering maps $\pi_1(G) \to \Z_n \hookrightarrow \U(1)$, with the trivial inclusion $\Z_n \hookrightarrow\U(1)$. As before, $\hom(\pi_1(G), \Z_n)\simeq \hom(H_2(BG),\Z_n)\simeq H^2(BG;\Z_n)$, so projective representations of Lie groups often reduce to central extensions by cyclic groups. This is for example the case of $\SO(3)$ whose projective representations are given by $H^2(B\SO(3);\Z_2)\simeq \Z_2$, the integer and half-integer spins (and indeed $H^2(B\SO;\U(1))\simeq \hom(\pi_1(\SO),\U(1))= \hom(\Z_2,\U(1))\simeq \Z_2$). This explains also the term 'fractionalization': projective representations are usually characterized by a non-trivial representation of $\Z_n\to \U(1)$, i.e. a map $m\to  e^{2\pi i m/n}$, with fractional charge $1/n$.}
Therefore, by a slightly abuse of language in the general case, we can say that after gauging $A$ in the extension \eqref{general grp ext}, the Wilson lines generating the dual symmetry $\wh A^{(d-2)}$ are fractionalized with respect to the remaining zero-form global symmetry $K$, i.e. they are in projective representations of $K$. This is a general phenomenon that could involve line defects in QFT; because in this case they are in fact the generators of another symmetry, there is the mixed anomaly \eqref{anomaly sym frac is grp ext anom} \cite{tachikawaGaugeFiniteGroups,brennanJacobsonRoumpedakis2025SymFrac}. A common example is the extension $\Z_2 \to \spin(d)\to \SO(d)$, where the condition \eqref{surface dependence wilson line ext} is indeed equivalent to say that the lines of $a$ are in projective representations of $\SO(d)$ specified by $\a\in H^2(B\SO;\Z_2)$. If $\SO(d)$ is the spacetime Lorentz group (and therefore $\Z_2$ is actually fermion parity $\Z_2^f=(-1)^F$), the Wilson lines represent fermionic particles (this is the case of bosonization \cite{gaiottokapustinspinTQFT1,thorngren2020anomalies,CappelliVillaBosDual2025,BergCappelliVilla2026SC,3dbosonization2024,gaiottokapustinspinTQFT2}).

This discussion applies the same for higher dimensional defects and higher groups. After gauging a finite Abelian $\wh A^{(d-q-2)}$ symmetry extending $K^{(p)}$ in a higher group, the generators $W(\S_{d-q-1})$ of the dual symmetry $A^{(q)}$ are fractionalized with respect to the remaining $K^{(p)}$ symmetry in the sense of \eqref{surface dependence wilson line ext}, with a class $[\a]\in H^{d-q}(B^{p+1}K;A)$. This again can be interpreted as an anomaly for the $d-q-1$ dimensional theory living on $W$.

Consider again the case \eqref{mixed U1 G group} from this perspective. After gauging $\Z_q$, its Wilson lines are fractionalized with respect to $\wt\U(1)\times \wt G$, with anomaly \eqref{mixed anomaly U1 G Z}. After gauging also $\wt \U(1)$, the Wilson lines for $\U(1)$ inherit the fractionalization with respect to the remaining  global symmetry $\wt G$ because of \eqref{wilson line A Atilde S} (the gauge field of $\Z_q$ is part of the total $\U(1)$ gauge field). They are generically not topological, but if we restrict to flat connections, they generate a magnetic $\Z^{(d-2)}$ symmetry and the mixed anomaly \eqref{mixed magnetic anomaly} is indeed the one coming from \eqref{anomaly sym frac is grp ext anom}. This is again a confirmation that the flat gauging picture captures all the relevant topological properties.

\section*{Acknowledgments}
The author thanks Marcus Berg and Andrea Cappelli for collaboration in the early stages of this work and for comments on the draft. The author is also grateful to Riccardo Argurio, Thomas Bartsch, Francesco Bonechi, Matteo Dell'Acqua and Antonio Santaniello for useful discussions, and to Yuji Tachikawa for helpful comments.

\appendix

\section{Symmetry fractionalization}
\label{app sym frac}

This appendix contains some facts about symmetry fractionalization phrased in a language similar to the one used in the main text and it serves as a complementary discussion to Section \ref{sec symfrac}. This subject is well explained in \cite{brennanJacobsonRoumpedakis2025SymFrac}, with attention to the details and spelling out its subtleties. 

\subsection{Neutral extended defects}
Broadly speaking, given a theory $\T$ with some global symmetry $G^{(0)}=G$,   symmetry fractionalization refers to the fact that line operators (and more generally extended defects) could have fractional quantum numbers with respect to local operators. This is a general phenomenon that does not require these extended defects to be charged under higher-form symmetries \cite{brennanJacobsonRoumpedakis2025SymFrac,barkeshli2014symfrac}. The most familiar case regards Wilson lines in gauge theories. If a Wilson line $W(\g)$ is not protected by any symmetry, it is endable and in particular there should be a gauge non-invariant local operator $O(x)$ on which the line could end. The symmetry $G$ acts on $O(x)$, which induces an action on $W(\g)$ itself \cite{brennanJacobsonRoumpedakis2025SymFrac}, which therefore should be in the same representation of $O(x)$ under $G$. However, there are two facts to keep in mind: first, $O(x)$ can be in a projective representation of $G$, being not gauge invariant;\footnote{A common case is if $\T$ has the symmetry structure 
\[ \frac{G_{gauge}\times G}{\G},\] where $\G \subseteq Z(G_{gauge})$. In such case the global symmetry acting on gauge invariant local operators is $G/\G$, while gauge non-invariant local operators that represent non-trivially $\G$ are in linear representations of $G$, therefore projective representations of $G/\G$.} second, given another local gauge invariant operator $\phi(x)$, $(\phi O)(x)$ (suitable regularized) is again an admissible endpoint for the line $W(\g)$. But $\phi(x)$ is generically in a linear representation of $G$, so the only meaningful statement one can make about the representation of the Wilson line itself is its projective class modulo tensoring with linear representations, classified by $H^2(BG;\U(1))$ \cite{brennanJacobsonRoumpedakis2025SymFrac,DumitrescuCordovaBrennanSymFrac2022}. It is possible to interpret this fact by saying that the quantum mechanical degrees of freedom living on the line (representing some higher energy particles integrated out) carry a 't Hooft anomaly for $G$: by anomaly matching, this should protect the line in the IR (i.e. it could not become the trivial line) \cite{brennanJacobsonRoumpedakis2025SymFrac}. 

What just said implies an action of the zero-form symmetry on extended objects, which is expected. For example, in a theory with charge conjugation symmetry, this acts on the Wilson lines of theory by permuting them (sending the one with charge $q$ to the one with charge $-q$). In general, a $p$-form symmetry acts on all operators of dimension greater then $p$ and this action should be taken into account. While the action of the disconnected symmetries is usually more obvious (like charge conjugation above), the one realized by continuous symmetries could be more subtle. In the following it is always assumed that the label of an extended defect is not changed by a lower dimensional symmetry.\footnote{So, in a theory with symmetry $G$, its action on the set $S$ of lines, given by $\rho: G \to \aut(S)$, is trivial. It can be taken into account \cite{KomargodskiHsinSymFrac2022}.}

By generalizing the above discussion on line operators, one can say that an extended defect $W(\S_p)$ supported on a submanifold $\S_p$ of dimension $p$ is fractionalized with respect to a global symmetry $G$ if the theory living on $\S_p$ carries an anomaly for $G$, which is given by $H^{p+1}(BG;\U(1))$. This can be also argued as follows. The action of $G^{(0)}$ on a Wilson line can be understood by simply noticing that the junction of two zero-form symmetry defects is a codimension $2$ locus, which therefore acts naturally on the line via linking.\footnote{The fact that this codimension $2$ submanifold is inevitably attached to higher dimensional symmetry defects implies that the line must pierce such manifolds somewhere. This is what makes this action sensible to local counterterms and the result is that only the projective class of the representation of the line under $G^{(0)}$ is scheme-independent \cite{DumitrescuCordovaBrennanSymFrac2022,brennanJacobsonRoumpedakis2025SymFrac}.} To act on a $p$-dimensional defect, $p+1$ zero-form symmetry defects should meet to give a $(d-p-1)$-dimensional junction that could link $W(\S_p)$ giving a phase: this yields $H^{p+1}(BG;\U(1))$. Even more generally, the fractionalization of $W(\S_p)$  with respect to a $q$-form symmetry $G^{(q)}$ is given by $H^{p+1}(B^{q+1}G;\U(1))$.

\subsection{With higher-form symmetries}

The discussion about symmetry fractionalization can be enriched when there are higher-form symmetries \cite{DumitrescuCordovaBrennanSymFrac2022,KomargodskiHsinSymFrac2022,brennanJacobsonRoumpedakis2025SymFrac,HsinShao2020,HsinTurzillo2019SETQuantumSpinLiquids}.

\subsubsection{Topological defects}

This is the case treated in the main text in Section \ref{sec symfrac}.  If the defects fractionalized with respect to some symmetry $G^{(p)}$ are themselves generators of another symmetry, there is the mixed anomaly \eqref{group extension anomaly app}.

Notice that in this case the fractionalization class can be reduced by consistency conditions \cite{brennanJacobsonRoumpedakis2025SymFrac}. For example, consider a topological line operator $W(\g)$ which generates a $\Z_n^{(d-2)}$ symmetry. Since $W^n=1$, the phase  $\a$ given by the action of $G^{(0)}$ symmetry defects on it must satisfy $\a^n=1$. Therefore, the fractionalization class is reduced from $H^2(BG;\U(1))$ to $H^2(BG;\Z_n)$.

\subsubsection{Charged defects}

Consider the case when the line, or the extended defect, is charged under a higher-form symmetry \cite{DumitrescuCordovaBrennanSymFrac2022,KomargodskiHsinSymFrac2022}.

Recall that the junction of two zero-form symmetry defects acts on the line by linking. When there is a one-form symmetry $G^{(1)}$ under which the line is charged, one could consider to enrich this junction with another symmetry defect for the one-form symmetry, which gives a phase when acting on the line, shifting its projective representation under $G^{(0)}$. Formally, given two symmetry defects of $G^{(0)}$, there is a symmetry defect for $G^{(1)}$, defined by a map $\th:G^{(0)}\times G^{(0)}\to G^{(1)}$. Consistency under associativity requires that $\th$ is a cocycle. This kind of mixing between $G^{(0)}$ and $G^{(1)}$ is thus controlled by an element $[\th] \in H^2(BG^{(0)};G^{(1)})$. In terms of the (flat, if the groups are continuous) dual background fields, $B_{(2)}=A_{(1)}^*\th$, with $A_{(1)}:X \to BG^{(0)}$.

More generally, we could extend this discussion to arbitrary two higher-form symmetries $A^{(q)}$ and $G^{(p)}$. Here the notation is valid when the groups are finite and Abelian (similar to \eqref{gauge field grp ext higher form}), but everything applies also to continuous groups if one restricts to flat background gauge fields (and $G$ can be non-Abelian when $p=0$). The two symmetries can mix non-trivially to form a group extension or a higher group, given by \eqref{gauge field grp ext higher form}. In such a case, they are not actually independent symmetries and indeed their gauge transformations mix them \eqref{gauge field grp ext higher form}: $G^{(p)}$ is a source for $A^{(q)}$. If the class \eqref{alpha as cohomol op} vanishes, then we say that the higher group splits and we have two independent symmetries. However, we could now consider a class $[\th] \in H^{q+1}(B^{p+1}G;A)$, equivalent to say that at the junction of $q-p+1$ $(d-p-1)$-dimensional $G$ defects there is a $(d-q-1)$-dimensional defect for $A$. In terms of the dual background gauge fields, this is 
\begin{equation}
    \label{sym frac gauge fields}
    A = G^* \th =\th(G), \qquad [\th] \in H^{q+1}(B^{p+1}G;A).
\end{equation}
The second expression is for the finite group case. We call such $\th$ a choice of symmetry fractionalization. It follows that $q \geq p$ for this procedure to be sensible (when $q=p$, every symmetry defect of $G$ comes with a symmetry defect of $A$).

Notice that the relation \eqref{sym frac gauge fields} does not mix the two symmetries, which are still good symmetries on their own. Indeed, even if $\th$ depends in some way on the representative of $G$, we have $\th(G+\d \l)= \th(G)+\d\g$, for some $\g \in C^{q}(X;A)$, which can be absorbed in a standard gauge transformation for $A$. Moreover, we see from \eqref{sym frac gauge fields} that $A$ is still closed even if $G$ is non-zero, different from \eqref{gauge field grp ext higher form}. However, as noticed in \cite{KomargodskiHsinSymFrac2022,DumitrescuCordovaBrennanSymFrac2022}, the relation \eqref{sym frac gauge fields} may affect 't Hooft anomalies for $G^{(p)}$, when there are $A^{(q)}$ anomalies or mixed $A-G$ anomalies, and it is thus important to consider cases like \eqref{sym frac gauge fields}: it could be necessary for UV-IR anomaly matching. 

As said, it is always assumed a trivial action $\rho: G\to \aut(A)$. This can be taken into account by considering a twisted cohomology group in \eqref{sym frac gauge fields}, as in \cite{KomargodskiHsinSymFrac2022}. More generally, in presence of more then two symmetries, one should consider the possible symmetry fractionalizations of a $q$-form symmetry for all $p<q$ symmetries. 

Notice that if $A^{(q)}$ and $G^{(p)}$ fit into a higher group extension \eqref{grp ext higher-form} specified by $[\a]\in H^{q+2}(B^{p+1}G;A)$, the relation \eqref{sym frac gauge fields} is not really meaningful. As remarked in the main text, the constraint \eqref{gauge field grp ext higher form} is not enough on its own and the whole pair $(A,G)$ must be given from the start. In particular, any closed part of $A$ is already encoded in the data of the pair, therefore \eqref{sym frac gauge fields} does not contain any extra physical information. By itself, \eqref{sym frac gauge fields} is not enough, since it does not satisfy \eqref{gauge field grp ext higher form}. This is the known fact that a higher group structure prevents to talk consistently about symmetry fractionalization (it is sometimes called an \textit{obstruction to symmetry fractionalization}) \cite{tachikawaGaugeFiniteGroups,CordovaBeniniHsin2group}.\\

\noindent\textit{Observation.} The expression `symmetry fractionalization' always refers to the fractional quantum numbers carried by extended operators, which is a general feature of such objects, being them charged or not under a higher symmetry. The most common case in $d=3,4$ regards line operators. When these objects are charged, one could consider also a possible additional symmetry fractionalization class in $H^2(BG^{(0)};G^{(1)})$, as in \eqref{sym frac gauge fields}, which gives a phase that stacks on the previous effect. In general $H^2(BG^{(0)};\U(1))\neq H^2(BG^{(0)};G^{(1)})$ and even when $G^{(0)}$ has no projective representations (e.g. $\Z_n$) there can be a non-trivial action given by \eqref{sym frac gauge fields}. In particular, it is this latter data that could modify the 't Hooft anomalies of $G^{(0)}$, not the general fact that a line carries a projective representation of it. This situation could be confusing because in some cases $H^2(BG^{(0)};\U(1)) = H^2(BG^{(0)};G^{(1)})$,
but one should keep in mind that there two different effects \cite{KomargodskiHsinSymFrac2022,brennanJacobsonRoumpedakis2025SymFrac} (one could say that the first one regarding neutral lines is more like a dynamical effect, while the second one \eqref{sym frac gauge fields} is just a necessary specification of the symmetry structure of the theory).\\

\subsubsection{Charged topological defects}

The two previous effects can be stacked. Consider the simple example of the $\Z_2$ gauge theory $\int_X b \cup (\d a + w_2)$. On one hand, this is level 2 BF theory, which is a bosonic theory with spacetime symmetry $\SO(d)$, where the $\Z_2^{(1)}$ symmetry (with background $B$) is fractionalized with respect to $\SO(d)$ according to \eqref{sym frac gauge fields} given by $B= w_2(TX)$ (fractionalization of spacetime symmetries will be discussed again later, but the bottom line is just that they are as the other symmetries). This shifts the projective representation of the Wilson line of $a$ which becomes a fermion \cite{thorngren2014framed}. On the other hand, the Wilson line of $a$ generates a $\Z_2^{(d-2)}$ symmetry, with background $C$, which therefore has the mixed anomaly  \eqref{group extension anomaly app} with Lorentz $\SO(d)$, here $i\pi\int_Y w_2\cup C$. From the first point of view, the BF mixed anomaly of $\Z_2^{(1)}$ with $\Z_2^{(d-2)}$, i.e. $i\pi \int B\cup C$, implies the mixed $\SO(d)$-$\Z_2^{(d-2)}$ anomaly since $B=w_2$. From the second point of view, this is indeed the anomaly expected given that the topological lines of $\Z_2^{(d-2)}$ carry a projective representation of $\SO(d)$. Notice that this is the general result of gauging the fermion parity symmetry $(-1)^F$ of a fermionic theory, namely bosonization, and indeed the resulting bosonic theory should have the aforementioned anomaly \cite{gaiottokapustinspinTQFT1,thorngren2020anomalies,CappelliVillaBosDual2025}. The BF theory considered as an example here can be interpreted as the result of gauging $\Z_2^f$ in the trivial fermionic TQFT \cite{CappelliVillaBosDual2025,gaiottokulp2020orbifolds,BergCappelliVilla2026SC}.

\subsection{Spacetime symmetries and Spin$_c$}

The discussion about symmetry fractionalization applies also to spacetime symmetries \cite{KomargodskiHsinSymFrac2022,HsinShao2020,DumitrescuCordovaBrennanSymFrac2022}. Generically, an extended defect could carry a projective representation of a spacetime symmetry and \eqref{sym frac gauge fields} could be extended to the case where $G$ is a spacetime background gauge field (in the connected part).

Consider the case of line operators. In a fermionic theory, where there are local half-integer spin particles and the spacetime group is $\spin(d)$, there is no notion of fractionalization since $H^2(B\spin(d);A)=0$ for every Abelian $A$ ($d\geq 3$).\footnote{
For a connected and simply connected group $G$, $H^2(BG;A)=0$ for every Abelian group $A$.} In a bosonic theory instead, where the spacetime group is $\SO(d)$, line operators could carry a projective representation of $\SO(d)$: lines in a linear representation of $\SO$ represent bosonic particles, while lines in projective representations represent fermionic particles (notice that in a bosonic theory there are no local fermionic operators, but this does not prevent to have extended objects with fermionic statistics). Physically, this means that in a bosonic theory we can consistently assign a spin modulo 1 to each line defect, while in a fermionic theory this assignment is consistent only modulo $1/2$: one can always dress the line with the fundamental neutral fermion, by changing its spin \cite{seibergFQHE2025}. If the lines are also charged under a one-form symmetry, there is also the fractionalization class \eqref{sym frac gauge fields} to take into account.

As an example, consider free Maxwell theory, i.e. bosonic $\U(1)$ gauge theory with spacetime symmetry $\SO(4)$, so $d=4$. The theory has two classes of line defects, Wilson lines $W(\g)$ and 't Hooft lines $H(\g)$, charged under the electric and magnetic one-form symmetries $\U(1)_e^{(1)}\times \U(1)_m^{(1)}$. In a standard treatment these lines are regarded as bosonic lines in a linear representation of $\SO(4)$, representing bosonic particles. However, given they are charged, one could consider the effect of \eqref{sym frac gauge fields}. What matters here is $[\th =w_2] \in H^2(B\SO;\Z_2)$, so, if $\G: X \to \SO(4)$ is  the background gauge field for $\SO(4)$,
\begin{equation}
\label{sym frac Maxwell}
    B_e= \G^* w_2=w_2(TX),\qquad B_m = \G^*w_2=w_2(TX).
\end{equation}
Starting from bosonic lines, the first choice corresponds to shift the projective representation class of the Wilson line $W$, making it fermionic, while the second choice make the 't Hooft line $H$ fermionic. Considering both of \eqref{sym frac Maxwell} makes all the lines in the theory fermionic, $W$, $H$ and the dyon $WH$. This is sometimes called all-fermion electrodynamics and it has a gravitational anomaly $w_2 w_3$ \cite{DumitrescuCordovaBrennanSymFrac2022, wittenNewSU2Anomaly2018, thorngren2014framed}: this comes from the standard electromagnetic mixed anomaly \cite{gaiotto2015generalized} with the fractionalization choice \eqref{sym frac Maxwell}. Notice that all this discussion requires the existence of just a $\Z_2$ subgroup of the one-form symmetries of the free theory, so it applies also to bosonic theories with $\U(1)$ gauge group where at least a $\Z_2$ subgroup of the electric or magnetic symmetry survives.

The first choice in \eqref{sym frac Maxwell}, $B_e=\pi w_2$ in a continuous notation, is equivalent to say the we consider the dynamical gauge field to be a $\spin_c$ connection \cite{DumitrescuCordovaBrennanSymFrac2022,BergCappelliVilla2026SC}. Indeed, when $B_e\neq 0$, the gauge invariant combination is $\d a- B_e$. $B_e=\pi w_2$ does not modify the local dynamics of $a$, but one could introduce a new connection $\wt a$ such that $\d \wt a=\d a -B_e= \d a-\pi w_2$. The fluxes of $\wt a$ are quantized as 
\begin{equation}
    \int \frac{\d \wt a}{2\pi} =\int\frac{\d a}{2\pi} +\frac{1}{2}w_2 = \Z+ \int \frac{1}{2}w_2,
\end{equation}
which is indeed the defining property of a spin$_c$ connection. This is consistent with the Wilson line being a fermion.\footnote{With the notation here, \[e^{i\oint \wt a}= e^{i\oint_\g a-i\pi\int_\S w_2},\] which shows that the Wilson line of the spin$_c$ connection $\wt a$ is well-defined but not a genuine line operator if $X$ is not spin. The genuine line operator is \[e^{i\oint_\g \wt a+i\pi\int_\S w_2}.\] However, this object transforms under electric one-form symmetry $B_e \to B_e +\d \l$, that with the identification $B_e =\pi w_2$ implies $w_2 \to w_2+ \frac{1}{\pi}\d \l$. When $w_2 =\d \eta$ this is equivalent to shift the spin structure $\eta \to \eta +\frac{1}{\pi}\l$, showing that the Wilson line is a fermion. The fact that the Wilson line of a dynamical spin$_c$ connection is not a genuine line operator is consistent with the fact that it represents a fermion in a bosonic theory where $(-1)^F$ is gauged, so that a fermionic particle is always attached with a line to be well-defined (gauge invariant).} By same reasoning, the choice $B_m=\pi w_2$ is equivalent to say that the dual connection is spin$_c$ and therefore the fundamental monopole represented by the 't Hooft line is a fermionic particle. Such choice implies a coupling $w_2 \d a /2\pi$ in the action, which, as observed in \cite{thorngren2014framed,wittenNewSU2Anomaly2018}, indeed makes the monopole a fermion.

Notice that the discussion regarding the fractionalization of the electric one-form symmetry is independent from the spacetime dimension. On the other hand, the magnetic symmetry is in general a $(d-3)$-form symmetry, so the discussion must be modified accordingly. For example, in $d=3$, the magnetic symmetry is a standard zero-form symmetry and the relevant group in \eqref{sym frac gauge fields} is $H^1(B\SO;\Z_2)=0$, so there are no possible fractionalization choices. This is coherent with the fact that the charged operator under $\U(1)_m^{(0)}$ is a local operator and therefore it cannot carry a projective representation of the global symmetry, here $\SO$. In a bosonic theory this operator must be bosonic and one cannot shift its projective representation with \eqref{sym frac gauge fields}, since making this operator a fermion would require the spacetime symmetry to be $\spin(3)$ for the theory to be consistent. In $d=5$, the magnetic symmetry is a 2-form symmetry and $H^3(B\SO;Z_2)=\Z_2[w_3]$. A fractionalization choice $B_m=\pi w_3$ would imply a 't Hooft surface describing the worldsheet of a fermionic string \cite{thorngren2014framed}.

\newpage
\bibliographystyle{ieeetr}
\bibliography{bibliography.bib}

\end{document}